# How do managers' non-responses during earnings calls affect analyst forecasts---a measurement by large language models

**Qingwen Liang***

Gies College of Business

University of Illinois Urbana-Champaign

Illinois, USA

Email: lqingwen@outlook.com

**Matias Carrasco Kind**

Gies College of Business

University of Illinois Urbana-Champaign

Illinois, USA

Email: mcarras2@illinois.edu

*Dr. Qingwen Liang is the corresponding author. Email: qingwen@illinois.edu. Postal address: 705 W. Main St. Urbana, IL, USA.

# How do managers' non-responses during earnings calls affect analyst forecasts—a measurement by large language models

## ABSTRACT

This paper examines the impact of managers' non-responses (NORs) during quarterly earnings calls on analyst forecast behavior by developing a novel measure of NORs using two mainstream large language models (LLMs): Chat-GPT4 and Llama 3.3. We adopt a three-step prompting approach–namely identification, classification, and evaluation–to extract NORs from the earnings call transcripts of S&P 500 firms. We document significantly positive relations between NORs and analyst forecast errors, dispersion, and uncertainty. We hypothesis that the positive relation is driven by greater information asymmetry and uncertainty. Consistent with the assumption, the cross-sectional analysis shows that the positive relationship between NORs and analysts' forecast features is concentrated in the set of sample firms operating across multiple industries, with higher institutional ownership and R&D expenditures, and holding earnings calls during the COVID period. In further analysis, we find that earnings calls with more NORs are followed by greater post-earnings announcement drift, indicating that NORs increase information asymmetry by raising information processing costs. Additionally, we find that NORs are associated with more return volatility, higher trading volume and wider bid-ask spreads, suggesting that NORs aggravate information uncertainty. Overall, our study demonstrates managers' non-responses during earnings calls exacerbate the information environment of analyst forecasts.

## 1. Introduction

Among the various channels of information disclosure, earnings conference calls have been attracting growing attention from researchers in accounting and finance. Typically, an earnings call consists of two parts: the managerial presentation and Q&A sections. Q&A sections are distinctive in providing "two-way" interactions where investors, analysts, and other participants can ask a company's management for further information regarding their presented performance. This interactive format provides a unique opportunity to observe managers' disclosure strategies in real-time communications (e.g., Bushee and Huang [2024]). Although the Q&A portion in earnings call is originally designed to promote transparency of public firms[1], in practice, managers often avoid or decline to answer questions (hereinafter referred to as "managers' non-responses" (NORs)). Despite their prevalence, the underlying motivations and potential consequences of managerial non-responses during earnings calls remain underexplored. On the one hand, managers may withhold bad news due to career concerns or personal wealth(e.g., Bao et al. [2019]). On the other hand, a large number of papers have shown that managers with good intentions would also reduce disclosures (e.g., Huang and Liang [2024]). Existing studies have documented a negative market reaction to managers' non-responses in earnings calls, indicating that investors perceive NORs as indicative of forthcoming bad news (e.g., Hollander, Pronk and Roelofsen [2010], Gow, Larcker and Zakolyukina [2021], Bian, Chen and Wang [2022], Barth, Mansouri and Woebbeking [2023]). Their conclusions are primarily based on the premise that investors are unable to distinguish between well-intended managerial non-responses and those aimed at concealing negative information. This sumption may not hold when it comes to different information recipients. Analysts constitute the primary participants during the Q&A portion of the earnings calls. Through their interactions with firm managers, analysts can obtain additional information to revise their forecasts. Distinct from retail investors, analysts are professional information processors and direct participants of earnings calls. When facing managers' non-responses, analysts are likely in a better position to discern the underlying motivations than common investors. In this paper, we examine the implications of managers' non-responses during quarterly earnings calls through the lens of analyst forecasts—an important stakeholder overlooked by prior research.

A key challenge to investigate this question is identifying the instances of managers' non-responses during earnings calls on a large scale. Previous research has primarily employed manual annotations and lexicon-based approaches to capture managers' non-responses within the call transcripts (e.g., Hollander, Pronk and Roelofsen [2010], Gow, Larcker and Zakolyukina [2021], Bian, Chen and Wang [2022]). Nevertheless, these methods are limited in their ability to detect managers' non-responses accurately and efficiently due to the inherent ambiguity and nuanced nature of human language. Recent advancements in large language models (LLMs), such as GPT-4, may provide a a better alternative to solve this problem. With its advanced ability in understanding and comprehension, LLMs offer distinct advantages in processing large-scale textual data (e.g., Kok [2025]). For example, human beings are often constrained by their limited attention and influenced by emotions, leading to unstable working outcomes. While LLMs, on the other hand, are able to maintain consistent performance by processing information systematically without emotional interruptions[2]. Compared to traditional machine learning approaches (ML) or natural language processing tools (NLP), LLMs can quantify textual information by integrating the full context (e.g., Siano [2025]), consequently making them better a batter option for text analysis. Leveraging their pre-training on massive textual datasets from diverse sources, LLMs are capable of detecting more implicit forms of non-responses, such as blathering or rambling. In addition, LLMs are more intuitive to start with, as they eliminate the need for extensive coding typically required in most ML or NLP approaches. Given the aforementioned advantages, we posit that LLMs can serve as a powerful tool for the efficient and effective identification of managerial non-disclosures in earnings call transcripts.

---

1. Regulation Fair Disclosure (Regulation FD) enacted in 2000 mandates that U.S. listed firms should make their earnings calls accessible to the public.URL:https://www.sec.gov/news/testimony/051701wssec.htm

2. In a recent experiment, Wang and Wang [2025] find LLMs significantly outperform expert human annotators in consistency and maintain high agreement even where human experts significantly disagree.

In this study, we develop a novel measure of managers' non-responses during earnings conference calls by experimenting on the most representative open-source (Llama3.3) and close-source LLMs (Chat-GPT4), respectively, aiming to tackle the following questions: 1. Can LLMs understand human languages and capture managers' non-responses (NORs) during an earnings call? 2. With the new measure, we are going to examine how the behavior of financial analysts relates to managers' non-responses.

ChatGPT4 and Llama3.3, like other large language models, perform downstream tasks via prompting[3]. Therefore, we adopt a three-step prompting approach–namely, identification, classification, and evaluation–to capture and construct the measure for managers' non-responses (NORs). Our original sample includes 10,976 copies of earnings call transcripts of S&P500 companies with 107,564 episodes of manager-analyst Q&A exchanges ranging from 2019Q1 through 2024Q3. A conference call usually starts with management presentations through which managers provide prepared interpretations of the firm's performance for this quarter, followed by Q&A sections where participants[4] can request further details by asking questions. Our analysis breaks down the Q&A sections and utilizes each consecutive conversation between an analyst and the management as the research unit. Using a manually coded dataset of 200 Q&A sections as the benchmark, we developed a few-shot prompt involving three tasks. Task 1 (Identification) instructs ChatGPT to identify managers' non-responses[5]. To gain deeper insights into how managers justify their inability to respond to the question in a conversation, task 2 (Classification) further requires ChatGPT4 to classify the non-responses into five distinct categories including "Refusal", "Lack of Info", "Legal Affairs", "Recall", and "Irrelevant". Task 3 (Evaluation)[6] instructs ChatGPT4 to evaluate managers' answers on a scale of 0-10 following three effective communication guidelines from Gricean Maxims (e.g., Grice [1989]) including the quantity of new information (Quantity), the degree of relevance(Relevance) and clarity (Clarity) of managers' responses to analysts' questions. With the identifications from ChatGPT4, we construct measures of managers' non-responses both on the conversation level ($NOR_C$ & $NOR^{Con}$) and the call level ($NOR_F$ & $NOR^{Firm}$).

We perform a battery of analyses to validate the appropriateness of the measures for man-agers' non-responses. Initially, we require ChatGPT4 to provide excerpts from the conversation unit to substantiate its identifications[7]. We manually checked 200 copies of its responses to make sure the excerpts are relevant to NORs. In the next, we prompt Llama3.3 with the same prompt and parameters. We find that, on average, Chat-GPT4 identifies approximately 23.3% more conversations with NORs than Llama3.3. However, statistical analysis indicates a similar year-trend and distribution across the two models.

---

3. In this procedure, all relevant task specifications and data process is formatted as textual input context, and the model returns a generated text completion.

4. Direct participants of an earnings call usually include institutional investors, analysts and managers.

5. In the prompt, we explicitly define non-responses as managers' refusals to answer despite of various reasons, irrelevant answers, or claims to call back in the future.

6. The purpose of Task 3 is to cross-validate with Task 1 to ensure the consistency of LLMs' outputs. Typically, scores from Task 3 are expected to be lower when managers provide a non-response.

We conduct examinations on the stability of LLMs' performances on the identifications of NORs through bootstrapping and regeneration processing. After performing bootstrapping with 100,000 iterations in each quarter, we portrays the mean values of NOR ratios over time. It demonstrates systematically consistent time trends in the identifications of NORs across the two models. We select 50 (38) conversations with NORs and 50 (62) conversations without any NORs by Llama3.3 (Chat-GPT4) to constitute the sample for regenerations. After repeating the prompting process 100 times, we calculate the matching ratios of two models. We find Chat-GPT4 (Matching Ratio: 76.22%) exhibit higher stability than Llama3.3 (Matching Ratio: 69.11%) in its identifications of NORs.

We regress Chat-GPT4-identified and Llama3.3-identified NORs on their evaluations about managers responses, features of Q&A sections, and other firm-level characteristics through both Logit model and OLS. The results from both models indicate that questions with a more negative tone, greater forward-looking content, and fewer requests for uncertainty-related information are more likely to elicit managerial non-responses.[8] These findings are largely consistent with those of Barth, Mansouri and Woebbeking [2023]'s. In addition, LLMs' evaluations of the quantity of incremental information and the relevance of responses to the questions are negatively related to managers' tendency to provide a non-response, indicating that Chat-GPT4 and Llama3.3 are internally coherent in their performances of the textual analysis task.

We test the relationship between the managers' non-responses and several properties of analyst earnings forecasts at the call level, including forecast errors, dispersion, and uncertainty. Theoretically, NORs may exert two opposing effects on the characteristics of analyst forecasts. On the one hand, NORs convey a signal that managers are inclined to withhold bad information regarding firms' earnings performances from the public, which could be observed and captured by analysts during the earnings call (*signal effect*). The bad signal itself, acting as a form of new information, can help analysts revise their earnings forecasts downward, improving precisions as well as reducing uncertainties. On the other hand, NORs, representing a lack of incremental information, can increase analysts' information asymmetry and intensify uncertainty about firms' fundamentals (*confusion effect*) [9].

Using LLMs' identified NORs for estimation, we find the "*confusion effect*" dominate the relationship between NORs and analyst forecast features. We document positive relationships between the occurrences of NORs and analyst forecast errors, dispersion and forecast uncertainty after controlling firm and quarter fixed effects. The main results are robust across a battery of tests including using different models and alternative variables, conducting analysis on individual analysts, as well as controlling managerial compensation incentives. We conduct an analysis using Chat-GPT4's evaluations of managers' answers based on three aspects of Gricean Maxims. The average value of *Quantity*, *Relevance*, and *Clarity* is negatively related to the characteristics of analyst forecasts, which is in line with the results of baseline regression. However, a more granular analysis reveals that only the score of *Relevance* is negatively related to analyst forecast features. Moreover, we examine the heterogeneous effects of distinct categories of NORs based on Chat-GPT4's classifications. Only *Direct Refusals* is positively associated with analyst forecast features, which implies that *Direct Refusals* primarily drives the impact of NORs on analysts' forecasts.

---

7. This process can also stand by ChatGPT4's advantages over traditional NLP tools in terms of the interpretability of its results.

8. Regression analyses using NORs from Chat-GPT4 and Llama 3.3 yield consistent results.

9. Following Zhang [2006], information uncertainty is defined as the ambiguity with respect to the implications of new information on a firm's value.

To substantiate our hypothesis, we analyze the impact of NORs on analysts' earnings forecasts in the contexts where analysts encounter varying levels of information asymmetry or uncertainty. Firstly, firms operating across multiple industries typically have more opaque information environments, creating obstacles for analysts to release accurate forecasts (e.g., Barinov, Park and Yıldızhan [2024]). Institutional investors with their professionalism can help disseminate useful information, leading to higher information transparency for those invested firms (e.g., Boone and White [2015]). We find that the significantly positive relationship between NORs and analyst forecast features is concentrated in the set of sample firms which operate in multiple industries or are owned by fewer institutional shareholdings, implying NORs affect analyst forecasts through exacerbating information asymmetry. Secondly, firms engaging in more intensive R&D activities suffer greater operational uncertainty, which could exacerbate the information uncertainty faced by analysts (e.g., Shi [2003], Kothari, Laguerre and Leone [2002]). During COVID-19, firms were faced with greater operational or financial uncertainties, thereby are subject to greater variations in their disclosures. We thus use R&D expenditure and COVID-19 period as measures for information uncertainty. The results show that the positive relationship between NORs and analyst forecast features is concentrated in the set of sample firms with higher R&D expenditures and earnings call held during COVID-19 period. Collectively, our evidence demonstrates that NORs affect analyst forecast features through increasing information asymmetry and intensifying information uncertainty.

We argue that NORs increase information asymmetry by rising information processing costs. To confirm the speculation, we examine the impact of NORs on post-earnings announcement drift (PEAD), a well-documented measure for information processing costs (e.g., Francis et al. [2007]). The results show that NORs amplify the PEAD effect for up to 60 days after the earnings announcement, demonstrating that NORs exacerbate information asymmetry by increasing information processing costs. Moreover, we examine the impact of NORs on the information uncertainty, proxy by stock price volatility, the average of trading volume and bid-ask spread in the 60 days following the earnings call. The results indicate that a higher frequency of NORs in earnings calls is associated with increased stock price volatility, higher trading volume, and a wider bid-ask spread, suggesting NORs amplify information uncertainty.

This paper contributes to the literature that investigates the consequences of managers' non-responses during a conference call. We create a novel measure for managers' non-responses in earnings calls by utilizing the most state-of-art large langue models. Despite the importance and ubiquity of managers' withholding information, measurement challenges make it difficult to evaluate its effect in the capital market. Specifically constrained by the inherent intricacies of human languages, quantitatively capturing managers' non-responses remains a hard endeavor. Three precedent papers undertook preliminary explorations regarding this problem. Hollander, Pronk and Roelofsen [2010] investigate whether managers withhold information from the public by manually reviewing and labeling the call scripts of publicly listed U.S. firms between Jan 1 and Dec 31 of 2004. Gow, Larcker and Zakolyukina [2021] develop the measures of non-responses by crafting a set of regular functions. Barth, Mansouri and Woebbeking [2023] identified 1364 trigrams which indicates non-answers by utilizing a supervised machine learning framework on a large training set of questions and answers from firms' earnings calls. These papers typically adopt rule-based methods to quantify textual data, which is primarily effective at identifying non-responses that are explicit and straightforward, but are limited in detecting more subtle or euphemistic occasions where managers evade analysts' questions. LLMs possess great flexibility in understanding human languages, making them a perfect tool to construct measures of interest based on contexts.

Theoretically, we extend prior studies assuming that managers' non-responses signal unfavorable information (e.g., Hollander, Pronk and Roelofsen [2010], Gow, Larcker and Zakolyukina [2021], Barth, Mansouri and Woebbeking [2023]). In contrast, we hypothesize and demonstrate that NORs with multiple motivations can actually confuse analysts, increasing their information uncertainty and asymmetry rather than providing an explicit informative signal. In response to Hollander, Pronk and Roelofsen [2010]'s call, our study also provides new evidence highlighting earnings calls' lack of credibility and liability, examined through the lens of analyst forecasts. To some extent, our findings echo those of Liu, Cao and Flake [2024], who claim that managers' non-responses to analysts' questions during an earnings call are driven by a mix of underlying incentives.

On top of that, this paper adds to the burgeoning literature regarding the applications of AI, NLP techniques, and large language models in the fields of accounting and finance (e.g., Kim, Muhn and Nikolaev [2024], Kok [2025]). By introducing LLMs' evaluations of managers' answers from three aspects of Gricean Norms—Clarity, Relevance, and Quantity of new information—our method offers a new approach on addressing the complexities and ambiguities of textual data. Our study provide a reference to future research that utilizes large language models to conduct textual analysis. Beyond simply identifying instances of non-response, we additionally classify them into five distinct categories, enabling deeper insights into both the features of these non-responses and their underlying causes. ChatGPT4 facilitates a granular analysis of managers' non-responses, allowing us to link different types of non-responses to their perceived outcomes.

## 2. Literature Review

### 2.1 LARGE LANGUAGE MODELS AND TEXTUAL ANALYSIS

The advent of large language models, with their advanced abilities in understanding, summarizing, and generating texts, provides a new solution to extract useful information from unstructured textual data. Several studies have explored the potential of using large language models as sophisticated research assistants for processing textual data in the fields of accounting and finance. Kim, Muhn and Nikolaev [2023a] apply the GPT 3.5 model to generate risk summaries from conference call transcripts, demonstrating that GPT-based measures offer significant information content, outperforming existing risk measures in predicting firm-level volatility and strategic choices. Kim, Muhn and Nikolaev [2023b] leverage ChatGPT's ability to summarize MD&A and earnings conference calls to develop a novel measure of information "bloat," offering a direct assessment of information processing costs. Similarly, Zang, Zheng and Zheng [2024] use GPT-2 and BERT after pre-training and fine-tuning, to develop a new measure of language predictability score (LPS) of MD&A by imitating the language ability of investors. They find that LPS outperforms the fog index, the bog index, and file size in explaining analysts' processing costs. Armstrong [2023] uses the fine-tuned ChatGPT model to create new measures of tax enforcement based on 10-Ks. He finds the model can actively identify ongoing IRS audits at a 96% accuracy rate compared to a tax researcher manually labeling the same disclosure. Bai et al. [2023] introduce a novel measure of information content by exploiting the discrepancy between answers to questions at earnings calls provided by corporate executives and those given by several context-preserving Large Language Models (LLM) such as ChatGPT, Google Bard, and an open source LLM. Yang [2023] leverages ChatGPT to design a predictive model for patent value. He finds LLM embeddings can capture many qualitative aspects of textual information previously unavailable to researchers. Hansen and Kazinnik [2023] evaluate both the zero-shot and fine-tuned ChatGPT-3 model to classify the policy stance of Federal Open Market Committee announcements. Their results show that ChatGPT models obtain the lowest numerical errors, the highest accuracy, and the highest measure of agreement relative to human classification compared with alternative NLP methods. Lopez-Lira and Tang [2023] document that ChatGPT can predict price movements using news headlines without direct financial training, suggesting the great potential of AI systems in benefiting information diffusion and decision-making. Kim, Muhn and Nikolaev [2024]'s research shows that LLMs exhibit a relative advantage over human analysts in their ability to predict earnings changes directionally.

These pioneering papers conduct preliminary explorations into the application of large language models (LLMs) to assist research in the accounting and finance domains. They demonstrate that LLMs have significant potential in understanding large-scale textual data and summarizing abstract concepts for further quantitative or empirical analysis.

## 2.2 MANAGERS NON-RESPONSES DURING EARNINGS CALLS

Existing research has identified several factors contributing to managers' strategic disclosures, including proprietary costs (e.g., Verrecchia [1983], Berger and Hann [2007]), poor earnings performance (e.g., Graham, Harvey and Rajgopal [2005], Bloomfield [2002], Kothari, Shu and Wysocki [2009]), self-interest (e.g., Cheng, Luo and Yue [2013]), information uncertainty (e.g., Dye [1985], Jung and Kwon [1988]), career concerns (e.g., Nagar, Nanda and Wysocki [2003], Armstrong, Guay and Weber [2010]), and litigation risks (e.g., Skinner [1997], Field, Lowry and Shu [2005], Houston et al. [2019]). However, these studies are limited in capturing managers' deliberate withholding of information (e.g., Hollander, Pronk and Roelofsen [2010]). Chen, Matsumoto and Rajgopal [2011] use firms' giving-up of earnings guidance as a proxy for the management's deliberate holding of information. They find that firms stop guidance because of poor prior performance, increased uncertainty and decreased informed investors. Hollander, Pronk and Roelofsen [2010] manually reviewed and labeled 1194 copies of earnings call transcripts with managers' non-responses. Their results show that firm size, CEO's stock_based incentives, company age and performance, litigation risks and whether analysts are actively involved during the call's Q&A section are best predictors of managers' non-responses. Moreover, they find investors interpret silence negatively. Gow, Larcker and Zakolyukina [2021] constructs a novel measure of managers' non-responses with linguistic analysis. They mainly investigate the motivations of managers' non-responses from earnings performance and proprietary costs separately, but remains quite limited on the consequence of managers' non-responses. Barth, Mansouri and Woebbeking [2023] identify 1364 trigrams signaling non-answers in earnings calls with a supervised machine learning framework on a large training set of questions and answers. Their measures are significantly associated with lower cumulative abnormal stock return and higher implied volatility. The most recent paper pertinent to our research is a literature review by Kok [2025], in which he performs a case study using ChatGPT recognize non-answers. His case provides a comprehensive description of non-answers identified by ChatGPT. However, his prompts are different from ours in many ways. For instance, he takes a one-shot strategy without any examples, while we attach two distinct examples for Chat-GPT to better understand tasks. Moreover, he uses individual question-and-answer pairs as the unit of analysis, whereas we make the best use of ChatGPT's comprehension capabilities by providing it with the complete back-and-forth dialogues between analysts and managers, ensuring no implicit information around the context is overlooked.

Overall, previous research is limited in their explorations regarding managers' non-responses,

with a focus on the their motivations. Although analysts are the primary questioners during conference calls, the impact of non-responses (NORs) on their forecasting behavior remains largely underexplored.

## 3. Theoretical Development

As professional information intermediaries in the capital market, analysts play a pivotal role in the discovery, interpretation, and dissemination of firm information. The Q&A sections of earnings conference calls offer analysts an opportunity to actively interact with firms' management, requesting additional information. Given that analysts are active participants of earnings calls, managerial non-responses to analysts' questions can have a significant effect on their forecasting behaviors (e.g., Bowen, Davis and Matsumoto [2002], Matsumoto, Pronk and Roelofsen [2011]). Previous studies shows that analysts acquire information through their interactions with the management during the earnings calls to update their earnings forecasts (e.g., Kimbrough [2005], Bushee, Gow and Taylor [2018]). Unlike the disclosure scenarios where outsiders are uninformed of and thus have no expectations for managers' private information, managers are presented with explicit information demands during a earnings call. Their refusals, deferrals, dodging, or obfuscations in response to analysts' questions are publicly observable. The managerial non-response is not only a proxy for reduced information but also a signal indicating managers' withholding of information. Consequently, we posit managerial non-responses during an earnings call can result in outcomes in two directions: 1. the *confusion effect* of reduced disclosures; 2. the *signal effect* from the very act of NORs. The question of concern is: From the perspective of analyst forecasts, which effect of NORs dominates the other? We answer this questions by examining the effect of NORs on three properties of analysts' forecasts–forecast error, forecast dispersion and forecast uncertainty.

Firstly, the adverse selection theory in voluntary disclosures contends that firms possessing favorable information are committed to more disclosures in order to differentiate themselves from those with unfavorable information (e.g., Grossman and Hart [1980], Dye [1985], Diamond and Verrecchia [1991]). Consistent with this theory, a number of papers find that firms are more inclined to reducing disclosures if there is forthcoming bad news. Kasznik and Lev [1995] find that managers are more likely to issue guidance before earnings disappointments than before positive earnings surprises. Kothari, Shu and Wysocki [2009]'s research suggests that managers on average delay the release of bad news to investors. Bao et al. [2019] validate their conclusion by using the residual short interest as a proxy for managers' private bad news. Bian, Chen and Wang [2022] show that the market reacts less favorably after managers dodge the questions from investors in Q&A sessions of the earning conference calls, and irrelevant answers predicts unsatisfactory financial performance. Recent studies examining stock market reactions to analyst-manager interactions assume and demonstrate that managers' reluctance to answer analysts' questions during public earnings calls signals their intention to conceal un-favorable news (e.g., Hollander, Pronk and Roelofsen [2010], Gow, Larcker and Zakolyukina [2021], Barth, Mansouri and Woebbeking [2023]). In other words, managerial non-responses during earnings conference calls imply unfavorable information regarding firms' future performances. If this assumption holds, more NORs should lead to smaller forecast errors, lower forecast dispersion, and less forecast uncertainty (*signal effect*) as this signal is captured and incorporated by those professional information intermediaries.

However, as prior research suggests, managers' disclosure decisions are shaped by complex and multifaceted motivations (e.g., Dye [1985], Verrecchia [1983], Lang and Lundholm [1993]). For instance, numerous studies have demonstrated that managers withhold information to safeguard firms' competitive priorities (e.g., Verrecchia [1983], Healy and Palepu [2001], Ellis, Fee and Thomas [2012], Li, Lin and Zhang [2018], Huang and Liang [2024]). Consequently, a manager's reluctance to answer analysts' questions does not necessarily signal negative news. Instead, it may steam from disclosing strategies with good intentions or simply due to the lack of information. Thus, the heterogeneity of potential incentives behind NORs can increase the uncertainty that analysts face when forecasting firms' future earnings. As a consequence, we predict higher incidence of NORs will yield higher forecast uncertainty and greater dispersions among analysts (*confusion effect*).

In addition, NORs can also substantially increase the information asymmetry. Managers who are unwilling to disclose information but concerned about the potential negative consequences of NORs may intentionally increase the complexity of their responses (e.g., Aghamolla and Smith [2023], Kim, Wang and Zhang [2019], Lo, Ramos and Rogo [2017]), obfuscating and confusing the audience. Jiang et al. [2024] take the online interactions as the research scenario. They find that firms that give less relevant answers to investors' questions tend to have larger analyst forecast bias and higher analyst optimism. Their results show that question dodging in firm-investor interactions can exacerbate firms' information asymmetry. Highly motivated by the reputation concerns, analysts are active in interacting with the management and attainting information from an earnings call to facilitate their forecasts. Ambiguous responses or non-responses require analysts to devote more time and efforts to collect extra supporting information from alternative sources. Moreover, several studies provide evidence that analysts are subject to limited attention (e.g., Koester, Lundholm and Soliman [2016], Driskill, Kirk and Tucker [2020]). Failing to obtain answers from management makes it more difficult and costly for analysts to process other information delivered in the earnings calls in a timely manner, leading to higher forecast errors (*confusion effect*).

Analysts with their professionalism in processing firms' public disclosures and access to alternative private information (e.g., Li, Wong and Yu [2020]), will not simply interpret managers non-responses as unfavorable information. In addition, analysts are the direct interlocutors in the Q&A exchanges with the management, and as such, they are better positioned to interpret the implicit signals conveyed through NORs within the context of the conversa-tion (e.g., Bushee and Huang [2024])[10]. Given the multiple motivations behind NORs and variances in analysts' information processing capabilities, we argue that analysts will exhibit greater divergence in their interpretations of NORs. Accordingly, *confusion effect* predicts that analyst forecasts will be less accurate and have greater dispersion for quarterly earnings calls with more NORs.

Given the opposing consequences predicted by the *signal effect* and the *confusion effect*, we formulate our hypothesis in the null form:

**H1a:** Managers' non-responses have no effect on analyst forecast errors.

**H1b:** Managers' non-responses have no effect on analyst forecast dispersion.

**H1c:** Managers' non-responses have no effect on analyst forecast uncertainty.

## *4. Research design*

### 4.1 DATA AND SAMPLE

Our sample period starts from 2019Q1 and ends in the 2024Q3. Table 1 outlines our sample selection process. We download the quarterly earnings call transcripts of S&P500 firms from the S&P Global Capital-IQ database (initial sample includes 10,976 copies of earnings call transcripts with 107,564 episodes of Q&A exchanges)[11] through WRDS. We keep only the latest version if there are more than one entries of transcript ID for a firm and drop those transcripts attached to more than one company IDs. After acquiring the initial transcripts, we then extract the units for LLM processing. During an earnings call, the management would make presentations regarding the firms' performances of the current financial quarter, followed by the Q&A exchanges between the analysts and managers. In real-world communication, messages must be interpreted within their broader conversational context. Therefore, unlike prior studies that code individual Q&A pairs (e.g., Hollander, Pronk and Roelofsen [2010], Gow, Larcker and Zakolyukina [2021], Bian, Chen and Wang [2022], Barth, Mansouri and Woebbeking [2023]), we use the full sequence of back-and-forth exchanges between analysts and management (hereafter, Q&A exchanges) as the unit of analysis to avoid overlooking contextual information. We combine the transcript ID with the sequence of Q&A exchanges within each transcript to create a unique ID for each Q&A exchange (Converid), which allows us to analyze the features of NORs at the conversation level[12]. The extracted Q&A exchanges are further fed into Chat-GPT4 and Llama3 separately. We match company ID of transcripts with GVKEY of Compustat to obtain firms' quarterly fundamentals, and PERMNO of CRSP to acquire stock price, return, and trading volume. We next merge the datasets with I/B/E/S to obtain EPS forecasts, the number of analyst followings, and actual earnings. The overlap between transcripts and I/B/E/S consist of the sample for our main analyses. In addition, we attain institutional holdings from Thomson Reuters Institutional Holdings (13F form) database and information about firms' divisions from CRSP. After removing observations with missing control variables, our final sample contains 4,284 transcripts of quarterly conference calls from 268 firms, consisting of Q&A exchanges between analysts and managers.

---

10. Bushee and Huang [2024]' research shows that analysts can better incorporate informational signals conveyed in from managerial linguistic complexity than investors.

11. In Capital-IQ, we use keydeveventtypeid="48" to filter out earnings call transcripts.

12. The data structures of transcripts from Capital-IQ enable use to identify the order where one analyst-manager interaction ends and the next one begins.

4.2 VARIABLE CONSTRUCTION

*4.2.1. Managers' non-responses* We utilize Chat-GPT4 and Llama3.3 to analyze the earnings call transcripts. We obtain the access to ChatGPT-4 model via the API provided by Azure Cloud Service and OpenAI and deploy LLaMA3.3 locally on the clusters of Data Science Research Services (DSRS)[13] of Gies College of Business. One key challenge in leveraging LLMs for research is ensuring replicability, as the outputs generated can vary substantially with each instance of prompting. To mitigate this concern as much as possible, we set the 'temperature', 'frequency_penalty', and 'presence_penalty' parameters of the model to 0. Meanwhile, we set the max_tokens parameter to 5000, large enough to prevent truncated outputs in the Q&A pairs. After setting up and configuring the model, we manually coded 200 random Q&A sections[14]. Enlightened by Kok [2025], we employ a few-shot approach to formulate the prompt, which incudes instructions with two demonstrations. Using the manually coded 200 Q&A sections as a benchmark, we iteratively refine the prompt until the its overlap with labeled data surpasses 70%.

In the finally settled prompt[15], we explicitly instruct two models to perform three tasks sequentially, attached with two examples for guidance. **Task 1–Identification** requires LLMs to identify managers' non-responses and extract supporting texts. Based on the non-responses identified in Task 1, **Task 2–Classification** instructs LLMs to classify the non-responses into five categories including "*Refusal*", which applies when managers directly decline to answer a question without providing any justification; "*Lack of Info*", if managers justify their non-responses by the lack of information; "*Legal Affairs*", if managers can't respond due to legal barriers; "*Recall*", if managers explicitly indicate they will revisit this question sometime in the future ; and "*Irrelevant*", if the management rambled on with irrelevant nonsense to dodge the questions. We also ask LLMs to return "*Null*" if no such non-responses are identified in Task 1. As a validation of LLMs' responses for Task 1, **Task 3–Evaluation** instructs LLMs to evaluate managers' responses based on three aspects of Gricean norms: the quantity of new information

---

13. Website of DSRS: https://dsrs.illinois.edu/

14. We create a new column called NOR, which is labeled as 1 if any questions are left unanswered in the conversation and return 0 for otherwise. To be specific, we adopt two strategies to code the Q&A exchanges. The first approach is straightforward. We identify situations where managers directly refuse to answer questions, either with or without a justification. The second approach relies on research assistants' subjective interpretation to identify more subtle instances of managers' non-answers, such as incomplete or irrelevant responses.

15. See Appendix A

(*Quantity*), relevance (*Relevance*), and clarity (*Clarity*). Ratings are assigned on a scale of 0 to 10, with higher scores indicating more informative, relevant and clear responses separately. To make our results traceable and verifiable, we instruct LLMs to output its responses in JSON format with 6 keys including "NOR", "Category","Inform","Relevant", and "Clarity".

We construct measures for managers' non-response at both call and conversation level. We define an an indicator ($NOR_C$) that equals 1 if at least one non-response is recognized during a Q&A exchange and 0 otherwise. If more than one question from a Q&A pairs are detected as unanswered, Chat-GPT will return the number of managers' non-responses($NOR^{Con}$). We manually look into Q&A sections identified with more than one NORs for an additional check. Additionally, we compute two non-response measures at the call level. One measure, $NOR_F$, is a binary indicator of managers' non-responses at the call level. It equals 1 if any non-responses occur during a quarterly earnings call and 0 otherwise. The second measure($NOR^{Firm}$) quantifies the total number of managers' non-responses in an earnings call, calculated as the sum of non-responses across all conversations. We will use $NOR^{Firm}$ in most of the empirical regression and other measures for statistics and robustness check.

*4.2.2. Analyst Revisions* We focused on three features of analyst forecasts, including forecast error, precision and uncertainty. **Error** is calculated as the absolute difference between the I/B/E/S reported quarterly EPS and the most recently issued analyst consensus following the earnings announcement date, scaled by the closing price of the previous fiscal quarter. Equation (1) shows the calculation of forecast error. **Dispersion** is computed as the standard deviation of the individual analyst forecasts in the most recent analyst forecast summary in I/B/E/S issued after the earnings call, scaled by closing price of the previous fiscal quarter. Equation (2) shows the calculation of forecast dispersion.

$$Error_{it} = \frac{|Mean\ Value\ for\ EPS\ Forecasts_{it} - Actual\ Value\ for\ EPS\ Forecasts_{it}|}{Closing\ Price_{it-1}} \quad (1)$$

$$Dispersion_{it} = \frac{Standard\ Deviation\ of\ Analyst\ EPS\ Forecasts_{it}}{Closing\ Price_{it-1}} \quad (2)$$

Referring to Bozanic and Thevenot [2015], We define analyst overall uncertainty as the sum of the idiosyncratic uncertainty and common uncertainty. The idiosyncratic uncertainty refers to the uncertainty associated with analysts' private information, while common uncertainty reflects the uncertainty related to information common to all analysts[16]. **Uncertainty** is computed using equation(3):

$$Uncertainty_{it} = \left(1 - \frac{1}{Analyst\ Following_{it}}\right) * Dispersion_{it} + Error_{it} \quad (3)$$

---

16. Bozanic and Thevenot [2015] consider both idiosyncratic uncertainty and overall uncertainty. While theoretically, managers' NORs are publicly observable by all analysts, which are mostly reflected in the overall uncertainty. So our paper only considers analyst overall uncertainty.

*4.2.3. Control Variables*   We control a series of variables that have been demonstrated by previous literature to be associated with analyst forecast behaviors. Following Lehavy, Li and Merkley [2011] and Bowen, Davis and Matsumoto [2002], we control firms size (**Size**), market value (**Mkv**), and unexpected earnings(**UEPS**). We use the natural logarithm of total assets plus one as of the end of last quarter as a proxy for firm size. We compute market value as the natural logarithm of the product of common shares outstanding and the closing price at the end of the last quarter plus one. Unexpected earnings are calculated as the difference between the current EPS and the EPS from the previous quarter, divided by the stock price at the end of the last quarter. To control for firms' earnings performance, we include return on assets (**Roa**) and a binary variable indicating whether the firm has negative earnings(**Loss**). Return on assets is computed as the income before extraordinary items divided by total assets as of the quarter end. We also add return volatility(**RetVol**) and firms' leverage(**Lev**) to control operation risks and financial risks separately. Return volatility is measured as the standard deviation of monthly stock returns over the 12 months preceding the earnings announcement date. Leverage is the computed by the total debt divided by total assets. To exclude the effects of presentation parts in earnings calls, we control an array of textual features of presentations, including language tone(**Tone**), uncertain information disclosures(**Uncert**), forward-looking information disclosures(**Forward**), the readability(**Read**). We calculate the tone of managers' presentations by using the difference between the frequency of positive words and the frequency of negative words divided by the sum of them. The positive and negative word list comes from Loughran and McDonald [2011]. The uncertainty in information disclosure is measured as the ratio of the frequency of uncertainty-related words to the total word count in the presentation. The uncertain word list comes from Loughran and McDonald [2011], with emphasis on the general notion of imprecision. The forward looking information disclosure is computed as the ratio of the frequency of forward-looking words to the total word count in the presentation. We acquire forward-looking terms from Bozanic, Roulstone and Van Buskirk [2018]. Following Li [2008] and Loughran and McDonald [2014], we utilize fog index as the measure for readability of managers' speeches. Appendix B provides further detailed descriptions of the control variables.

### 4.3 MODEL DESIGN

Our baseline regression is to examine the effect of managers' non-responses (NOR$_{it}$) on the features of analysts' forecasts by estimating the following model:

$$Forecasts_{it} = \beta_0 + \beta_1 NOR_{it}^{Firm} + \beta_2 Controls_{it} + \sum Firm + \sum Quarter + \varepsilon_{it} \qquad (4)$$

Where *Forecasts* are the three features of analyst forecasts at the firm level which include Error$_{it}^{Firm}$, Dispersion$_{it}^{Firm}$, and Uncertainty$_{it}^{Firm}$. In robustness tests, we also estimate the regression at the analyst individual level, consisting features of individual analyst forecast errors (Error$_{it}^{Individual}$) and response time (Time$_{it}^{Individual}$). All regressions are estimated with firm fixed effects and quarter fixed effects. Continuous variables are winsorized at 1 percent in both tails of the distribution to remove the effects of outliers.

## 5. Results

### 5.1 DESCRIPTIONS OF NOR-RESPONSES

Before diving into the empirical analysis, we conduct descriptive statistics to have an overview of the main measures. We start with a comparison between Chat-GPT4 and Llama3.3.

*5.1.1. Descriptions of Non-responses: Chat-GPT4 VS. Llama3.3* Panel A of Table 1 presents the identifications (Task1) of managers' non-responses. We performed the tasks with both Chat-GPT4 and Llama3.3. Our sample consists 107,564 episodes of conversations between analysts and the management in total. Chat-GPT4 identified 15,645 conversations as containing non-responses (NORs) across all Q&A exchanges, including 15,384 conversations with one question receiving NOR, 257 with two questions receiving NORs, and 4 with three questions receiving NORs. We also find that 416 conversations are subject to output formatting problems with Chat-GPT4[17]. Llama3.3 identified 12,689 (12,682+7) conversations as having NORs, including 12,682 episodes with one question receiving NOR and seven episodes with two questions receiving NORs. There are 335 conversations that are not successfully processed by Llama3.3[18]. On average, Chat-GPT4 identified 23.3% more conversations with NORs than Llama3.3.

Panel B reports the classifications(Task2) of managers' non-responses. Chat-GPT4 classified 1,588 NORs as irrelevant answers, 6,895 NORs as lack of information, 401 NORs as legal concerns, 2,546 as direct refusals without justification, and 4,465 as future recall. In task2, Llama3.3 classified most NORs as lack of information (8,274) or direct refusals (2,749). It identified 1,872 NORs as irrelevant answers, 622 NORs as future recalls, and 121 NORs as legal concerns. As shown by the results, both models classified the majority of non-responses (NORs) as either a lack of relevant information or direct refusals. However, the models differ substantially in the *Recall* category, where Chat-GPT4 identified nearly seven times as many instances as Llama3.3. To sum up, Chat-GPT4's classifications about NORs tend to be more specific, with relatively fewer NORs classified as direct refusal, lack of information, and irrelevant answers than Llama3.3. In contrast, it classified more NORs with explicit reasons, such as legal constraints and intentions to revisit the question in the future.

---

17. To facilitate downstream processing, we instruct the language models to generate responses in JSON format. However, we observed the models occasionally produce outputs that do not adhere strictly to the specified format (Errors) during the experiment. We exclude Error occasions in the empirical analysis.

18. ChatGPT-4 reported more errors than Llama 3.3, which may be attributed to the differences in how the two models are accessed. Llama 3.3 is hosted locally on DSRS's internal clusters, whereas ChatGPT-4 is accessed via Azure's API.

Panel C and D present the statistics of Chat-GPT4's and Llama3.3's evaluations of managerial responses based on the three Gricean Rules. Both models assigned higher scores for Relevance and lower scores for Quantity in managers' responses, implying that while managers' answers are relevant to the questions, they are limited in offering incremental information. Once again, we observe more errors reported by Chat-GPT4.

**TABLE 1**
*Descriptions of Non-responses*

| Panel A: Numbers of Non-Responses | | | | | | |
|---|---|---|---|---|---|---|
| Types | NOR=0 | NOR=1 | NOR=2 | NOR=3 | NOR=Error | All |
| ChatGPT | 91,503 | 15,384 | 257 | 4 | 416 | 107,564 |
| Llama | 94,540 | 12,682 | 7 | 0 | 335 | 107,564 |
| Panel B: Types of Non-Responses | | | | | | |
| Types | Irrelevant | Lack | Legal | Refusal | Recall | Other |
| ChatGPT | 1,588 | 6,895 | 401 | 2,546 | 4,465 | 8 |
| Llama | 1,872 | 8,274 | 121 | 2,749 | 622 | 405 |
| Panel C: Evaluations of Managerial Responses from Chat-GPT4 | | | | | | |
| Dimensions | Mean | Sd. | Min | Median | Max | Count | Error |
| Quantity | 7.78 | 0.98 | 0.00 | 8.00 | 10.00 | 107,429 | 135 |
| Relevance | 9.24 | 1.23 | 0.00 | 10.00 | 10.00 | 107,429 | 135 |
| Clarity | 8.52 | 0.96 | 0.00 | 9.00 | 10.00 | 107,429 | 135 |
| Panel D: Evaluations of Managerial Responses from Llama3.3 | | | | | | |
| Dimensions | Mean | Sd. | Min | Median | Max | Count | Error |
| Quantity | 7.68 | 1.00 | 0.00 | 8.00 | 9.00 | 107,564 | 0 |
| Relevance | 8.81 | 0.80 | 0.00 | 9.00 | 10.00 | 107,564 | 0 |
| Clarity | 8.82 | 0.80 | 0.00 | 9.00 | 10.00 | 107,564 | 0 |

This table presents descriptive statistics of the responses from Chat-GPT4 and Llama3.3 respectively. Panel A reports the number of non-responses identified in a Q&A exchange by two models (Task1: Identifications). Panel B reports the classifications of non-response types by the two models (Task2: Classifications). Panel C and D present the statistics of LLMs' evaluations of managerial responses.

Figure 1 illustrates the overlap between Chat-GPT4-identified NORs and Llama3.3-identified NORs. Approximately 60.22% (74.23%) of Chat-GPT4 (Llama3.3) identified NORs overlap with Llama3.3's (Chat-GPT4's). The two models identified 9,422 NORs in common. Figure 2 presents the distribution of NORs at both the call and conversation levels. The number of NORs identified by Chat-GPT4 (Llama3.3) per earnings call ranges from 1 (1) to 13 (11). At the Q&A exchange level, the number of NORs identified by Chat-GPT4 (Llama3.3) ranges from 1 (1) to 3 (2).

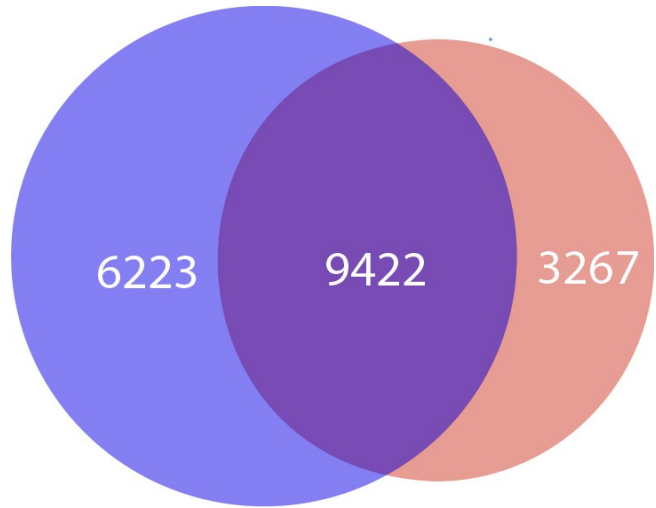

FIG. 1 — Venn Diagram of Non-responses.

This figure is the venn diagram capturing the overlap in NORs between Chat-GPT4 and Llama3.3, respectively. The circle in blue represents Chat-GPT4-identified NORs and the circle in red represents the Llama3.3-identified NORs. The overlapping part is NORs which are captured by both models.

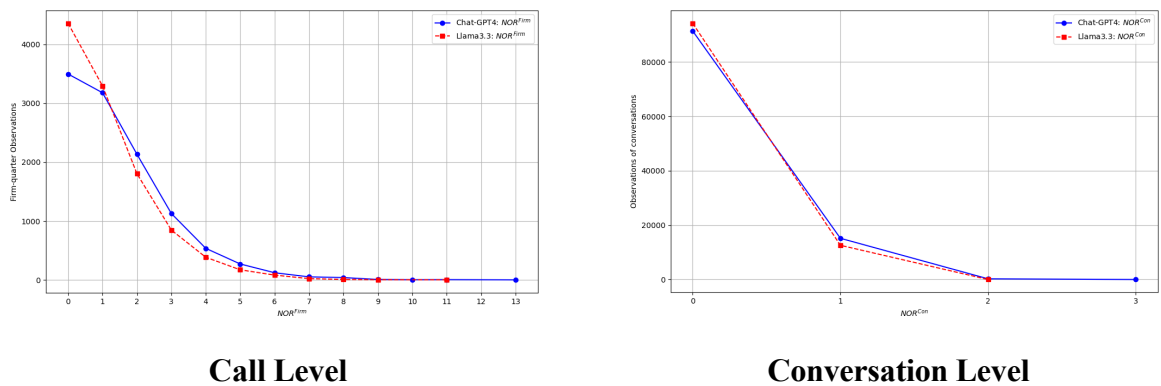

**Call Level**  **Conversation Level**

FIG. 2 Distributions of Non-responses

This figure presents the distributions of NORs on call level (the left) and on the conversation level (the right) respectively. The Y-axis represents the number of calls (conversations), and the X-axis represents the number of NORs identified in every call (conversation) unit. The line in blue indicates the results from Chat-GPT4. The line in red indicates the results from Llama3.3.

*5.1.2. Non-responses Statistics by Quarter* We calculate NOAs ratios at call and conversation level. The NOR ratio at call level ($NOR_F$) is computed as the percentage of calls with NORs to the total number of calls of each quarter. The NOR ratio at conversation level ($NOR_C$) is computed as the percentage of conversations with NORs to total number of conversations of each quarter. Table 1 in Online Appendix presents the yearly trend of $NOR_F$ and $NOR_C$ of Chat-GPT4(Panel A) and Llama3.3(Panel B). Column(2) shows that Chat-GPT4 identified 60.55% to 75.36% as having non-responses from 2019Q1 to 2024Q3 and Llama3.3 identified 50.26% to 66.26% earnings calls as having non-responses from 2019Q1 to 2024Q3. On the conversation level, Column(3) shows that Chat-GPT4 identified 12.02% to 18.27% Q&A exchanges as having non-responses from 2019Q1 to 2024Q3 and Llama3.3 identified 9.50% to 14.75% Q&A exchanges as having non-responses from 2019Q1 to 2024Q3. Column(4) through (8) provide a breakdown of $NOR_C$ into distinct categories based on LLMs' classifications in Task2.

*5.1.3. Time Series Analysis*  We further analyze the time-varying patterns of NORs ratios at the call and conversation level, respectively. Specifically, we perform bootstrapping with 100,000 iterations in each quarter and portrays the mean values of NOR ratios over time. The results are displayed in Online Appendix. Figure 1(Figure 3) in the Online Appendix show the call-level(conversation-level) NOR ratios of both models exhibit approximately normal distributions. We then plot the quarterly trend of NOR ratios of Chat-GPT4 (in blue) and Llama3.3 (in red) in Figure 2 (Call-level) and Figure 4 (Conversation-level). The solid lines represent the actual value of NOR ratios in each quarter and shaded areas are one standard errors above or below the mean value of NOR ratio. As it is shown, Chat-GPT 4 identified more NORs than Llama3.3 at both call-level and conversation level. However, their identifications exhibit systematically consistent time trend. Furthermore, the evolution of conversation-level NOR ratios depicted in Figure 4 closely mirrors the pattern observed in Figure 4 of Kok [2025], wherein the conversation serves as the primary unit of analysis.

*5.1.4. Assessment of the Stability of LLMs*  We validate the stability of LLMs' performances on the identifications of NORs. We start by randomly selecting 50 (38) conversations with NORs and 50 (62) conversations without any NORs by Llama3.3 (Chat-GPT4) to constitute the baseline sample. Then we repeat the prompting process for 100 times on both Chat-GPT4 and Llama3.3. For each conversation, match ratio is computed as the percentage of iterations in which a model (Chat-GPT or Llama) produced the same identification as the baseline sample. We report the match ratios in Table 2 of Online Appendix. On average, Chat-GPT4 tends to have higher match ratios (76.22% for Chat-GPT4 and 69.11% for Llama3.3) and relatively lower instability (38.04% standard deviation for Chat-GPT4 and 44.71% standard deviation for Llama3.3). This implies Chat-GPT4 produces more consistent and replicable results than Llama3.3. Notably, we find both models exhibit lower consistency in analyzing conversations without non-responses(NOR=0) compared to those with non-responses(NOR=1).

*5.1.5. Motivations for Non-responses*  We conduct two tests to validate our measures by examining the motivations of non-responses. Firstly, we cluster NORs with 24 potential influencing factors concluded by prior research, and the results are shown in Figure 5 of Online Appendix. We use UMAP to reduce the 24 factors into 2 dimensions, representing the X-axis and Y-axis separately. For both Chat-GPT4 and Llama3.3, the 24 factors effectively distinguish between conversations with NORs and those without, validating our selection of influencing factors for NORs. Secondly, we use OLS and logit model to regress NORs on their influencing factors. The results are reported in Table 3 of Online Appendix, the evaluations of quantity and relevance by both models are negatively associated with the probability of non-responses, indicating the models' evaluation standards are consistent. However, we find opposite results for Chat-GPT4 and Llama3.3 in terms of "*Clarity*". The results also show that questions asked later in the call, those with a negative tone, requests for more forward-looking information, and those seeking less uncertain information are more likely to receive non-answers. Additionally, Q&A sections with more words are negatively associated with NORs. The tests on the motivations of NORs justify our approach in capturing managers' non-responses in earnings calls.

## 5.2 VARIABLE STATISTICS

Panel A to D of Table 2 reports the summary statistics of the main variables used in this paper. All the continues variables are winsorized at the 1st and 99th percent. The number of NORs in every quarterly earnings call (NOR$^{Firm}$) ranges from 0 to 11 with a deviation of 1.522 in our sample. The number of NORs in each conversation (NOR$^{Con}$) ranges from 0 to 2 with a deviation of 0.358. On average, there are 67.5%(14.2%) earnings calls(Q&A exchanges) have non-responses. The mean is larger than the median for all analyst forecast features, suggesting a right skewness in their distributions. The mean and median of forecast errors on firm level (individual level) is 0.006 (0.577) and 0.002 (0.154), respectively. The mean and median of analyst forecast are 0.202 and 0.096, respectively. The mean and median of analyst forecast uncertainty are 0.176 and 0.087, respectively.

## 5.3 UNIVARIATE ANALYSIS

Panel E of Table 2 provides results of univariate analysis for the overall differences in analysts' forecasts and other variables of interest for earnings calls with or without NORs. Three features of analyst forecasts are all significantly (at the 1% level) higher for earnings calls with NORs, implying analyst forecast errors, dispersion, and uncertainty as a result of managers' non-responses in the earnings calls. Among the other control variables, RetVol, Loss, Mkv, Tone, and Read also show differences in two groups.

**TABLE 2**
*Variable Statistics and Univariate Analysis*

| Panel A: Managers' Non-responses | | | | | | |
|---|---|---|---|---|---|---|
| | N | Mean | Sd. | Min | 50% | Max. |
| $NOR^{Firm}$ | 4284 | 1.435 | 1.522 | 0.000 | 1.000 | 11.000 |
| $NOR_F$ | 4284 | 0.675 | 0.468 | 0.000 | 1.000 | 1.000 |
| $NOR^{Con}$ | 51513 | 0.144 | 0.358 | 0.000 | 0.000 | 2.000 |
| $NOR_C$ | 51513 | 0.142 | 0.349 | 0.000 | 0.000 | 1.000 |
| Panel B: Analyst Forecast Features | | | | | | |
| $Error^{Firm}$ | 4284 | 0.006 | 0.014 | 0.000 | 0.002 | 0.104 |
| Dispersion | 4284 | 0.202 | 0.309 | 0.002 | 0.096 | 2.063 |
| Uncertainty | 4284 | 0.176 | 0.265 | 0.003 | 0.087 | 1.787 |
| Panel C: Control Variables | | | | | | |
| SurEar | 4284 | 0.000 | 0.019 | -0.082 | 0.000 | 0.101 |
| Size | 4284 | 10.523 | 1.360 | 7.745 | 10.437 | 14.660 |
| Roa | 4284 | 0.015 | 0.018 | -0.045 | 0.013 | 0.076 |
| RetVol | 4284 | 9.010 | 3.553 | 3.518 | 8.210 | 21.662 |
| Loss | 4284 | 0.082 | 0.274 | 0.000 | 0.000 | 1.000 |
| Mkv | 4284 | 10.605 | 1.102 | 8.564 | 10.515 | 13.875 |
| Lev | 4284 | 0.292 | 0.160 | 0.005 | 0.287 | 0.715 |
| Tone | 4284 | 0.441 | 0.259 | -0.347 | 0.485 | 0.889 |
| Uncert | 4284 | 0.008 | 0.004 | 0.002 | 0.007 | 0.026 |
| Forward | 4284 | 0.017 | 0.006 | 0.004 | 0.016 | 0.035 |
| Read | 4284 | 10.246 | 1.182 | 6.660 | 10.220 | 13.070 |
| Panel D: Other Variables | | | | | | |
| H_RD | 4284 | 0.265 | 0.442 | 0.000 | 0.000 | 1.000 |
| Inst | 4284 | 0.560 | 0.496 | 0.000 | 1.000 | 1.000 |
| MO | 4262 | 0.567 | 0.496 | 0.000 | 1.000 | 1.000 |
| Ret_Sd | 4054 | 0.021 | 0.011 | 0.007 | 0.019 | 0.068 |
| Volume | 4054 | 11.088 | 1.067 | 8.917 | 10.984 | 14.030 |
| Spread | 4054 | 0.026 | 0.011 | 0.013 | 0.024 | 0.071 |
| $Time^{Individual}$ | 31987 | 0.567 | 0.652 | 0.000 | 0.693 | 2.639 |
| $Error^{Individual}$ | 31987 | 0.577 | 1.409 | 0.000 | 0.153 | 10.542 |
| CAR | 4092 | -0.003 | 0.124 | -0.810 | -0.002 | 0.545 |
| BHAR | 4092 | -0.005 | 0.124 | -0.631 | -0.007 | 0.677 |

| Panel E: Univariate Analysis | | | | |
|---|---|---|---|---|
| Variable | Mean ($NOR_F$=0) | Mean ($NOR_F$=1) | Mean-Diff | T-statistic |
| $Error^{Firm}$ | 0.005 | 0.007 | -0.001*** | -2.963 |
| $Dispersion^{Firm}$ | 0.182 | 0.211 | -0.029*** | -2.899 |
| $Uncertainty^{Firm}$ | 0.159 | 0.185 | -0.026*** | -3.043 |
| SurEar | 0.001 | 0.000 | 0.001 | 1.125 |
| Size | 10.490 | 10.538 | -0.049 | -1.099 |
| Roa | 0.016 | 0.015 | 0.001 | 1.164 |
| RetVol | 8.816 | 9.103 | -0.287** | -2.479 |
| Loss | 0.070 | 0.087 | -0.017* | -1.864 |
| Mkv | 10.645 | 10.585 | 0.060* | 1.679 |
| Lev | 0.289 | 0.294 | -0.005 | -0.986 |
| Tone | 0.478 | 0.423 | 0.055*** | 6.481 |
| Uncert | 0.008 | 0.008 | -0.000 | -0.276 |
| Forward | 0.017 | 0.017 | -0.000 | -0.706 |
| Read | 10.357 | 10.193 | 0.165*** | 4.281 |
| Observations | 1,391 | 2,893 | | |

Panel A to Panel D of the table present descriptive statistics for the main variables used in the regressions. All continuous variables are winsorized at the 1st and 99th percentiles. Panel E of the table presents the univariate differences in the mean of analyst forecast features and control variables for earnings calls conditional on whether having non-responses ($NOR_F$=0 VS. $NOR_F$=1). The last two columns provide the P-value and T-statistics from the T-test for the mean differences of two independent samples, respectively. *, ** and *** denote statistical significance at the 10%, 5% and 1% levels, respectively. All the variables are defined in Appendix B.

## 5.4 MAIN REGRESSION

Table 3 provides the empirical results of our main regression. We examine how the number of non-responses for each earnings call (NOR$^{Firm}$) affect features of analyst forecasts. Odd-numbered columns are the regression results without control variables and even-numbered columns represent regression results with control variables included. T-statistics, presented in brackets, are based on robust standards that are clustered at the firm level. The coefficient of (NOR$^{Firm}$) on analyst forecast error (Column(1)-(2)) and dispersion (Column(3)-(4)) are positive and statistically significant at 1% level, indicating that more NORs are associated with less accurate analyst earnings forecasts and more dispersed forecasts. In addition, we find that overall analyst uncertainty is increasing with the frequency of NORs, suggesting that there is higher overall uncertainty in the analyst information environment for firms with more NORs.

In terms of economic significance, a one-standard-deviation increase in NOR$^{Firm}$(1.435) increases analyst forecast errors by 0.0007 (1.435 × 0.00047)[19], which corresponds to about 11.24% of the average degree of analyst forecast errors. A one-standard-deviation increase in NOR$^{Firm}$(1.435) increases analyst forecast dispersion by 0.012 (1.435 × 0.008), corresponding to 5.7% of the average degree of analyst forecast dispersion. A one-standard-deviation increase in NOR$^{Firm}$(1.435) increases analyst forecast dispersion by 0.01 (1.435 × 0.007), corresponding to 5.7% of the average degree of the overall analyst forecast uncertainty.

Collectively, the results in Table 3 align with the prediction of *confusion effect*, suggesting NORs may exacerbate firms' information environments.

---

19. The coefficient of NOR$^{Firm}$ on Error$^{Firm}$ is 0.0004736, which was round to three decimal places as reported in Table 3

**TABLE 3**
*Managers' Non-responses and Analyst Forecasts*

|  | (1) Error$^{Firm}$ | (2) Error$^{Firm}$ | (3) Dispersion$^{Firm}$ | (4) Dispersion$^{Firm}$ | (5) Uncertainty$^{Firm}$ | (6) Uncertainty$^{Firm}$ |
|---|---|---|---|---|---|---|
| NOR$^{Firm}$ | 0.000*** | 0.000*** | 0.008*** | 0.008*** | 0.008*** | 0.007*** |
|  | (2.94) | (2.93) | (2.80) | (2.81) | (3.00) | (3.00) |
| SurEar |  | 0.015 |  | 0.916** |  | 0.784** |
|  |  | (0.71) |  | (2.40) |  | (2.43) |
| Size |  | 0.005*** |  | 0.131*** |  | 0.117*** |
|  |  | (3.08) |  | (4.29) |  | (4.44) |
| Roa |  | -0.056** |  | -0.734 |  | -0.630 |
|  |  | (-2.35) |  | (-1.57) |  | (-1.60) |
| RetVol |  | 0.000 |  | 0.006*** |  | 0.005*** |
|  |  | (0.53) |  | (2.93) |  | (3.13) |
| Loss |  | -0.001 |  | 0.012 |  | 0.010 |
|  |  | (-1.05) |  | (0.53) |  | (0.53) |
| Mkv |  | -0.006*** |  | -0.152*** |  | -0.135*** |
|  |  | (-4.92) |  | (-6.49) |  | (-6.68) |
| Lev |  | -0.002 |  | -0.046 |  | -0.036 |
|  |  | (-0.43) |  | (-0.50) |  | (-0.46) |
| Tone |  | 0.001 |  | -0.012 |  | -0.013 |
|  |  | (0.62) |  | (-0.64) |  | (-0.80) |
| Uncert |  | -0.008 |  | 1.314 |  | 0.997 |
|  |  | (-0.10) |  | (0.93) |  | (0.82) |
| Forward |  | 0.125** |  | 0.547 |  | 0.662 |
|  |  | (2.51) |  | (0.64) |  | (0.92) |
| Read |  | -0.000 |  | -0.007 |  | -0.007 |
|  |  | (-0.50) |  | (-1.19) |  | (-1.27) |
| Firm | Yes | Yes | Yes | Yes | Yes | Yes |
| Quarter | Yes | Yes | Yes | Yes | Yes | Yes |
| Constant | 0.006*** | 0.014 | 0.190*** | 0.451 | 0.166*** | 0.395 |
|  | (19.94) | (0.74) | (37.33) | (1.32) | (38.01) | (1.35) |
| Observations | 4284 | 4284 | 4284 | 4284 | 4284 | 4284 |
| Adj. $R^2$ | 0.308 | 0.319 | 0.526 | 0.546 | 0.532 | 0.554 |

This table presents the results of Eq. (5) accounting for the relationship between managers non-responses and the characteristics of managers' forecasts in terms of *Error*, *Dispersion* and *Uncertainty*. NOR$^{Firm}$ proxies for the number of NORs identified by Chat-GPT4 during the earnings call in quarter t of firm i. Columns (1), (3), and (5) show the regression results without control variables. In columns (2),(4), and (6), all the control variables are further included. The detailed definitions of the variables are provided in Appendix B. The t statistics, estimated on robust standard errors are reported in parentheses. *, ** and *** denote statistical significance at the 10%, 5% and 1% levels, respectively.

5.5 ANALYSIS OF NON-RESPONSE TYPES AND RESPONSE ASSESSMENT

*5.5.1. Classifications of Managers' Non-responses* Task 2 of the prompt instructs LLMs to classify managers' non-resnses into five categories including direct refusal (*Refusal*=1), irrelevant answers (*Irrelevant*=1), recall in the future(*Recall*=1), lack of necessary information (*Lack*=1), and legal concerns (*Legal*=1). Panel A of Table 4 shows the empirical results examining the heterogeneous effects of various categories of NORs on analyst forecast features. We find that managers' *direct refusal* dominates the relation between NORs and analyst forecast behaviors.

*5.5.2. Assessment of Managers' responses*  As a validation of Task 1, Task 3 in the prompt requires LLMs to evaluate the quantity and quality of managers' responses on a scale of 1 to 10 based on the three facets of Grice Rule. "*Quantity*" measures the quantity of incremental information in managers' responses. "*Clarity*" measures the level of precision and absence of ambiguity in managers' responses. "*Relevance*" refers to the extent to which managers' responses align with the analysts' questions. We first calculate the average of the three metrics and re-estimate the Equation (4). The results reported in Panel B of Table 4 show that Chat-GPT4's average evaluations in terms of the quantity and quality of managers' responses are negatively associated with analyst forecast errors, dispersion, and uncertainty, with statistical significance at the 1% level. Panel C of Table 4 presents the distinct effects of *Quantity*, *Clarity*, and *Relevance* on analyst forecast features, showing that only the score of *Relevance* is significantly and negatively associated with analyst forecast features. Overall, the results from the Chat-GTP4's assessments of managers' responses are in line with the findings from baseline regression, further validating the internal consistency of LLMs' performance in textual analysis.

**TABLE 4**
*Analysis of Non-response Types and Response Assessment*

| Panel A: Classifications of Non-responses | | | | | | |
|---|---|---|---|---|---|---|
| | (1) $\text{Error}^{Firm}$ | (2) $\text{Error}^{Firm}$ | (3) $\text{Dispersion}^{Firm}$ | (4) $\text{Dispersion}^{Firm}$ | (5) $\text{Uncertainty}^{Firm}$ | (6) $\text{Uncertainty}^{Firm}$ |
| Refusal | 0.001** | 0.001** | 0.014** | 0.014** | 0.013*** | 0.012** |
| | (2.54) | (2.48) | (2.52) | (2.46) | (2.60) | (2.53) |
| Irrelevant | 0.001* | 0.001* | 0.005 | 0.005 | 0.005 | 0.005 |
| | (1.73) | (1.74) | (0.61) | (0.66) | (0.79) | (0.85) |
| Recall | 0.000 | 0.000 | 0.010 | 0.010 | 0.009 | 0.009 |
| | (0.89) | (0.84) | (1.16) | (1.28) | (1.18) | (1.29) |
| Lack | 0.000 | 0.000 | 0.002 | 0.002 | 0.002 | 0.002 |
| | (0.36) | (0.40) | (0.56) | (0.55) | (0.66) | (0.65) |
| Legal | 0.000 | 0.000 | 0.027 | 0.023 | 0.024 | 0.020 |
| | (0.34) | (0.23) | (1.12) | (0.96) | (1.20) | (1.02) |
| Controls | No | Yes | No | Yes | No | Yes |
| Firm | Yes | Yes | Yes | Yes | Yes | Yes |
| Quarter | Yes | Yes | Yes | Yes | Yes | Yes |
| Constant | 0.006*** | 0.013 | 0.191*** | 0.440 | 0.166*** | 0.386 |
| | (20.16) | (0.70) | (37.17) | (1.29) | (38.00) | (1.32) |
| Observations | 4284 | 4284 | 4284 | 4284 | 4284 | 4284 |
| Adj. $R^2$ | 0.308 | 0.319 | 0.526 | 0.546 | 0.532 | 0.553 |

| Panel B: Average Evaluations of Managers' Responses | | | | | | |
|---|---|---|---|---|---|---|
| | (1) $\text{Error}^{Firm}$ | (2) $\text{Error}^{Firm}$ | (3) $\text{Dispersion}^{Firm}$ | (4) $\text{Dispersion}^{Firm}$ | (5) $\text{Uncertainty}^{Firm}$ | (6) $\text{Uncertainty}^{Firm}$ |
| Mscore | -0.002*** | -0.002*** | -0.036*** | -0.032*** | -0.032*** | -0.028*** |
| | (-3.03) | (-2.83) | (-3.13) | (-2.80) | (-3.21) | (-2.85) |
| Controls | No | Yes | No | Yes | No | Yes |
| Firm | Yes | Yes | Yes | Yes | Yes | Yes |
| Quarter | Yes | Yes | Yes | Yes | Yes | Yes |
| Constant | 0.023*** | 0.029 | 0.508*** | 0.716** | 0.447*** | 0.628** |
| | (4.17) | (1.52) | (5.17) | (2.04) | (5.29) | (2.09) |
| Observations | 4284 | 4284 | 4284 | 4284 | 4284 | 4284 |
| Adj. $R^2$ | 0.309 | 0.319 | 0.526 | 0.546 | 0.532 | 0.554 |

| Panel C: Granular Evaluations of Managers' Responses | | | | | | |
|---|---|---|---|---|---|---|
| | (1) $\text{Error}^{Firm}$ | (2) $\text{Error}^{Firm}$ | (3) $\text{Dispersion}^{Firm}$ | (4) $\text{Dispersion}^{Firm}$ | (5) $\text{Uncertainty}^{Firm}$ | (6) $\text{Uncertainty}^{Firm}$ |
| Relevance | -0.002* | -0.002* | -0.038** | -0.038** | -0.033*** | -0.033*** |
| | (-1.94) | (-1.89) | (-2.55) | (-2.57) | (-2.65) | (-2.67) |
| Quantity | -0.001 | -0.001 | -0.006 | -0.008 | -0.007 | -0.008 |
| | (-0.92) | (-1.05) | (-0.36) | (-0.48) | (-0.48) | (-0.62) |
| Clarity | 0.001 | 0.001 | 0.017 | 0.023 | 0.016 | 0.022 |
| | (1.03) | (1.23) | (1.11) | (1.53) | (1.20) | (1.65) |
| Controls | No | Yes | No | Yes | No | Yes |
| Firm | Yes | Yes | Yes | Yes | Yes | Yes |
| Quarter | Yes | Yes | Yes | Yes | Yes | Yes |
| Constant | 0.020*** | 0.027 | 0.450*** | 0.668* | 0.397*** | 0.587* |
| | (3.46) | (1.43) | (4.29) | (1.89) | (4.40) | (1.94) |
| Observations | 4284 | 4284 | 4284 | 4284 | 4284 | 4284 |
| Adj. $R^2$ | 0.309 | 0.319 | 0.527 | 0.547 | 0.533 | 0.554 |

Panel A of the table presents the empirical results examining the heterogeneous effects of various types of NORs on on analyst forecasts. The types of NORs come from Task2 which prompts Chat-GPT4 to classify identified NORs into five groups including *Refusal* for a direct refusal, *Irrelevant* for an irrelevant answer, *Recall* for a claim to feed back, *Lack* for lack of information, and *Legal* for legal limitations. Panel B of the table presents the empirical results examining the effect of the average value of Chat-GPT4's evaluations of managers responses on analyst forecast features. The evaluations are built on three aspects from Gricean Maxims: *Relevant*, *Quantity*, and *Clarity*. Panel C of the table reports the distinct effects of *Relevant*, *Quantity*, and *Clarity* on analyst forecast features. *Relevance* measures the degree to which managers' responses directly address analysts' questions. *Quantity* serves as a proxy for the incremental information conveyed in managers' answers. *Clarity* evaluates how clearly managers articulate their responses. For brevity, the coefficients of other control variables are omitted. The t statistics, estimated on robust standard errors are reported in parentheses. *, ** and *** denote statistical significance at the 10%, 5% and 1% levels, respectively.

## 5.6 ROBUSTNESS TEST

*5.6.1. Alternative Measures* We substantiate our results by testing Llma3.3-identified NORs on analyst forecast features. The results are reported in Panel A of Table 5. The coefficients of Llama_NOR$^{Firm}$ on forecast error, dispersion, and uncertainty are significantly positive, consistent with the Chat-GPT4's results.

We replace the independent variable with a dummy variable (NOR$_F$) that equals 1 if an earnings call contains any non-responses and 0 otherwise. The results are presented in Panel B of Table 5. NOR$_F$ is still positively related to features of analyst forecasts.

Following Lehavy, Li and Merkley [2011], we replace measure analyst forecast error by calculating the squared difference between the I/B/E/S reported earnings and the analyst consensus forecast, scaled by the closing stock price of last quarter (SquError$^{Firm}$). The Uncertainty is adjusted accordingly (SquUncertainty$^{Firm}$). The results are still consistent with the baseline regression, as reported in Panel C of Table 5.

*5.6.2. Analysis on Individual Analysts* Regressions on the summary of analyst forecasts can not capture individual analysts' reactions. To address this concern, we extend our analysis to examine the behavior of individual analysts directly. We acquire individual analyst EPS forecasts within 30 days after the earnings announcement day. Then we compute the absolute difference between analysts' forecasts and actual EPS divided by the stock price of last quarter as the measure for forecast errors (Error$^{Individual}$). Additionally, we regress on individual analysts' response time, which is computed as the algorithm of gap between the earnings announcement day and the most recent EPS estimation or revisions of individual analysts in 30 days plus 1. As reported in Panel D of Table 5, NOR$^{Firm}$ is positively associated with individual analyst forecast errors at 1% significant level, consistent with the baseline findings. We find a negative relation between NOR$^{Firm}$ and analyst response time, implying that analysts response more promptly to those earnings calls with NORs. Lehavy, Li and Merkley [2011]'s research shows that less unreadable reports would spur investors' demand for information interpretations, so firms with less readable financial reports are followed by more timely analyst forecasts. Building on their conclusion, we interpret a shorter response time as analysts' reaction to the market's demand for information processing following earnings calls with more NORs[20] Taken together, the analysis on individual analyst confirms that confusion effect dominates the relation between NORs and analysts forecast features.

*5.6.3. Controlling Managerial Incentives* Previous studies document that managerial stock incentive and wealth are important contributors to their disclosure strategies (e.g., Nagar, Nanda and Wysocki [2003], Hollander, Pronk and Roelofsen [2010]). Following Hollander, Pronk and Roelofsen [2010], we further control managerial stock-based compensa-tion (COMP) and stock price–based wealth (Lwealth). We identify the names of managers in each Q&A exchange and merge them with the managerial compensation from Execucomp database[21]. COMP is computed as the ratio of the average annual stock price–based compensation (i.e., the sum of the total value of stock-option grants plus the value of restricted stock grants) to total direct managerial compensation[22]. Lwealth is the natural logarithm of the value of managerial shareholdings in the firm (using the fiscal year-end closing price). If there are more than one managers participating in a Q&A exchange, we take the mean value of their COMP and Lwealth for our analysis. Panel E of Table 5 reports the results, NOR$^{Firm}$ remains positively related with features of analyst forecasts.

---

20. Since previous research reveals that information demand increase as uncertainty rises (Amiram et al. [2018]), this finding also echos the prediction of *confusion effect* by indicating an increase of information demand.

---

21. We use managers' names as the keyword to perform textual fuzzy match.
22. The total compensation in this paper includes bonus, salary, other annual cash awards, value of restricted stock grants, net value of stock option grants, long-term incentive payout, and all other annual compensation.

**TABLE 5**
*Robustness Tests*

| Panel A: Regression with LlaMa-identified NORs | | | | | | |
|---|---|---|---|---|---|---|
| | (1) | (2) | (3) | (4) | (5) | (6) |
| | Error$^{Firm}$ | Error$^{Firm}$ | Dispersion$^{Firm}$ | Dispersion$^{Firm}$ | Uncertainty$^{Firm}$ | Uncertainty$^{Firm}$ |
| Llama_NOR$^{Firm}$ | 0.000** | 0.000** | 0.008*** | 0.008*** | 0.008*** | 0.008*** |
| | (2.16) | (2.15) | (2.65) | (2.74) | (2.85) | (2.93) |
| Controls | No | Yes | No | Yes | No | Yes |
| Firm | Yes | Yes | Yes | Yes | Yes | Yes |
| Quarter | Yes | Yes | Yes | Yes | Yes | Yes |
| Constant | 0.006*** | 0.015 | 0.193*** | 0.469 | 0.168*** | 0.402 |
| | (20.85) | (0.80) | (40.11) | (1.38) | (40.92) | (1.38) |
| Observations | 4324 | 4324 | 4324 | 4324 | 4324 | 4324 |
| Adj. $R^2$ | 0.311 | 0.321 | 0.529 | 0.549 | 0.535 | 0.557 |

| Panel B: Dummy Variables for Non-responses | | | | | | |
|---|---|---|---|---|---|---|
| | (1) | (2) | (3) | (4) | (5) | (6) |
| | Error$^{Firm}$ | Error$^{Firm}$ | Dispersion$^{Firm}$ | Dispersion$^{Firm}$ | Uncertainty$^{Firm}$ | Uncertainty$^{Firm}$ |
| NOR$_F$ | 0.001** | 0.001** | 0.021*** | 0.020** | 0.018*** | 0.017*** |
| | (2.28) | (2.21) | (2.59) | (2.53) | (2.65) | (2.58) |
| Controls | No | Yes | No | Yes | No | Yes |
| Firm | Yes | Yes | Yes | Yes | Yes | Yes |
| Quarter | Yes | Yes | Yes | Yes | Yes | Yes |
| Constant | 0.006*** | 0.013 | 0.188*** | 0.444 | 0.164*** | 0.389 |
| | (16.69) | (0.72) | (30.88) | (1.30) | (31.34) | (1.33) |
| Observations | 4284 | 4284 | 4284 | 4284 | 4284 | 4284 |
| Adj. $R^2$ | 0.308 | 0.318 | 0.526 | 0.546 | 0.532 | 0.553 |

| Panel C: Alternative Measures for Analyst forecasts | | | | |
|---|---|---|---|---|
| | (1) | (2) | (3) | (4) |
| | SquError$^{Firm}$ | SquError$^{Firm}$ | SquUncertainty$^{Firm}$ | SquUncertainty$^{Firm}$ |
| NOR$^{Firm}$ | 0.038** | 0.037** | 0.079*** | 0.076*** |
| | (2.29) | (2.25) | (3.93) | (3.88) |
| Controls | No | Yes | No | Yes |
| Firm | Yes | Yes | Yes | Yes |
| Quarter | Yes | Yes | Yes | Yes |
| Constant | 0.012 | -0.030 | 0.169*** | -0.360 |
| | (1.44) | (-0.03) | (14.94) | (-0.35) |
| Observations | 4284 | 4284 | 4284 | 4284 |
| Adj. $R^2$ | 0.078 | 0.082 | 0.336 | 0.348 |

Panel A of the table presents the empirical results using LlaMa3.3-identified NORs to replicate the main regression. Llama_NOR$^{Firm}$ represents the number of NORs identified by Llama3.3 during the earnings call in quarter t of firm i. Panel B presents the empirical results using an alternative measure(NOR$_F$) for managers non-responses on the firm level to examine the effect of NORs on analyst forecast features. NOR$_F$ is a dummy variable which equals 1 if there are any non-responses during an earnings call and 0 otherwise. Panel C reports the empirical results using alternative measures of analyst forecast error (SquError$^{Firm}$) and uncertainty (SquUncertainty$^{Firm}$). Following Lehavy, Li and Merkley [2011], we measure analyst forecast error (referred as *Accuracy* in Lehavy, Li and Merkley [2011]'s paper) by calculating the squared difference between the I/B/E/S reported earnings and the analyst consensus forecast, scaled by the closing stock price of last quarter. SquUncertainty$^{Firm}$ is adapted following this change accordingly. For brevity, the coefficients of other control variables are omitted. The t statistics, estimated on robust standard errors are reported in parentheses. *, ** and *** denote statistical significance at the 10%, 5% and 1% levels, respectively.

**TABLE 5**
*Robustness Tests—Continued*

| Panel D: Analyst Forecasts–Individual Level Analysis | | | | |
|---|---|---|---|---|
| | (1) $Error^{Individual}$ | (2) $Error^{Individual}$ | (3) $Time^{Individual}$ | (4) $Time^{Individual}$ |
| $NOR^{Firm}$ | 0.037*** | 0.031*** | -0.006** | -0.007** |
| | (6.37) | (5.50) | (-2.10) | (-2.30) |
| Firm | No | Yes | No | Yes |
| Firm | Yes | Yes | Yes | Yes |
| Quarter | Yes | Yes | Yes | Yes |
| Constant | 0.523*** | -0.426 | 0.576*** | 0.441 |
| | (52.74) | (-0.52) | (107.13) | (1.46) |
| Observations | 31987 | 31987 | 31987 | 31987 |
| Adj. $R^2$ | 0.303 | 0.337 | 0.109 | 0.110 |

| Panel E: Managerial Incentives | | | |
|---|---|---|---|
| | (1) $Error^{Firm}$ | (2) $Dispersion^{Firm}$ | (3) $Uncertainty^{Firm}$ |
| $NOR^{Firm}$ | 0.000*** | 0.006* | 0.006* |
| | (2.65) | (1.66) | (1.82) |
| Comp | 0.004 | 0.179*** | 0.153*** |
| | (1.64) | (4.14) | (4.06) |
| Lwealth | 0.001** | 0.006 | 0.007 |
| | (2.04) | (0.64) | (0.97) |
| Controls | Yes | Yes | Yes |
| Firm | Yes | Yes | Yes |
| Quarter | Yes | Yes | Yes |
| Constant | 0.005 | -0.101 | -0.057 |
| | (0.16) | (-0.24) | (-0.16) |
| Observations | 2838 | 2838 | 2838 |
| Adj. $R^2$ | 0.361 | 0.600 | 0.611 |

Panel D presents the empirical results examining the effect of NORs on individual analysts' behaviors in terms of forecast errors(*Error*) and response time(*Time*). $Error^{Individual}$ represents the individual analyst's forecast errors, computed as the absolute difference between the EPS forecast of individual analyst j and the actual EPS for firm i in quarter t, divided by the closing price of quarter t-1. *Time* represents the individual analyst's response time, computed as the natural logarithm of the days between the earnings announcement day and the most recent EPS estimation or revisions of individual analysts in 30 days plus 1. Panel E shows the empirical result controlling the incentives of managers. Referring to Hollander, Pronk and Roelofsen [2010], we adopt two measures of managerial incentives that are tied to stock price: stock price–based compensation (COMP) and stock price–based wealth (Lwealth). COMP is calculated as the ratio of the average annual stock price–based compensation (i.e., the sum of the total value of stock-option grants plus the value of restricted stock grants) to total direct managerial compensation. Lwealth is the natural is the natural logarithm of the value of managerial shareholdings in the firm (using the fiscal year-end closing price). If there are more than one managers participating in the conversation, we take the average of their compensation to compute the two measures. For brevity, the coefficients of other control variables are omitted. The t statistics, estimated on robust standard errors are reported in parentheses. *, ** and *** denote statistical significance at the 10%, 5% and 1% levels, respectively.

### 5.7 HETEROGENEOUS TEST

To test our argument that more NORs increase analyst forecast errors and discrepancies by amplifying the information asymmetry and uncertainty, we partition our sample based on four variables: *MO* for firm complexity, *Inst* for sophisticated investor attention, *RD* for R&D investment amount, and COVID for the influence of COVID-19.

*5.7.1. Information Asymmetry* First, we posit that if information asymmetry is the channel through which NORs affect analyst forecasts, the results from our main regression should be more pronounced among firms with higher information asymmetry. Barinov, Park and Yıldızhan [2024] documents that firms with greater operational complexity exhibit higher levels of information processing costs and information asymmetry. Referring to their research, we define *MO* to measure firm operational complexity, which equals 1 if the number of a firm's divisions with different two-digit SIC codes are above the mean value and 0 if otherwise. We then split the sample into two groups based on the value of MO. We re-estimate Equation (4) in two sub-samples and the results are reported in Panel A of Table 6. The coefficients of NORs on forecast errors, dispersion and uncertainty are significantly positive (at 1% level) only in firm with higher operational complexity. We compare the coefficients on $NOR^{Firm}$ between sub-samples using Fisher's permutation test. As shown in the last line, the difference of the NORs coefficients between the two sub-samples is statistically significant.

**TABLE 6**
*Information Asymmetry*

Panel A: Firm Complexity

|  | Error$^{Firm}$ | | Dispersion$^{Firm}$ | | Uncertainty$^{Firm}$ | |
| --- | --- | --- | --- | --- | --- | --- |
|  | MO=1 | MO=0 | MO=1 | MO=0 | MO=1 | MO=0 |
|  | (1) | (2) | (3) | (4) | (5) | (6) |
| NOR$^{Firm}$ | 0.001*** | 0.000 | 0.013*** | 0.003 | 0.012*** | 0.003 |
|  | (2.69) | (1.34) | (2.85) | (0.93) | (3.03) | (1.03) |
| Controls | Yes | Yes | Yes | Yes | Yes | Yes |
| Firm | Yes | Yes | Yes | Yes | Yes | Yes |
| Quarter | Yes | Yes | Yes | Yes | Yes | Yes |
| Constant | 0.005 | 0.039* | 0.062 | 0.527 | 0.047 | 0.489 |
|  | (0.16) | (1.92) | (0.13) | (1.03) | (0.11) | (1.12) |
| Observations | 2416 | 1843 | 2416 | 1843 | 2416 | 1843 |
| Adj. $R^2$ | 0.327 | 0.298 | 0.508 | 0.615 | 0.509 | 0.637 |
| Diff. in coef. | P-Value=0.080* | | P-Value=0.002*** | | P-Value=0.002*** | |

Panel B: Institutional Ownership

|  | Error$^{Firm}$ | | Dispersion$^{Firm}$ | | Uncertainty$^{Firm}$ | |
| --- | --- | --- | --- | --- | --- | --- |
|  | Inst=1 | Inst=0 | Inst=1 | Inst=0 | Inst=1 | Inst=0 |
|  | (1) | (2) | (3) | (4) | (5) | (6) |
| NOR$^{Firm}$ | 0.000 | 0.001*** | 0.003 | 0.014*** | 0.002 | 0.013*** |
|  | (0.93) | (2.64) | (0.74) | (3.14) | (0.72) | (3.42) |
| Controls | Yes | Yes | Yes | Yes | Yes | Yes |
| Firm | Yes | Yes | Yes | Yes | Yes | Yes |
| Quarter | Yes | Yes | Yes | Yes | Yes | Yes |
| Constant | 0.005 | 0.006 | 1.185*** | -1.580** | 0.989*** | -1.312** |
|  | (0.26) | (0.15) | (4.01) | (-2.09) | (3.82) | (-2.05) |
| Observations | 2398 | 1875 | 2398 | 1875 | 2398 | 1875 |
| Adj. $R^2$ | 0.322 | 0.335 | 0.533 | 0.564 | 0.536 | 0.579 |
| Diff. in coef. | P-Value=0.047** | | P-Value=0.008*** | | P-Value=0.005*** | |

Panel A of the table compares the baseline regression results between firms which are more complicated in their operations (*MO*=1) and firms which are less complicated (*MO*=0). We measure firms' complexity by counting the number of firms' divisions with different two-digit SIC codes, where more divisions represent higher operating complexity. We report the p value of the Fisher's Permutation test (two-sided) for the difference in the coefficient on NOR$^{Firm}$ between the two sub-samples in the last row. Panel B of the table compares the baseline regression results between firms with more attention from sophisticated investors(*Inst*=1) and those with less attention from sophisticated investors(*Inst*=0). We measure sophisticated investor attention using institutional holdings, as institutional investors are typically regarded as sophisticated processors of firms' information. We report the p value of the Fisher's Permutation test (two-sided) for the difference in the coefficient on NOR$^{Firm}$ between the two sub-samples in the last row. For brevity, the coefficients of other control variables are omitted. The t statistics, estimated on robust standard errors are reported in parentheses. *, ** and *** denote statistical significance at the 10%, 5% and 1% levels, respectively.

Institutional investor holdings can serve as an external governance mechanism regulating firms' disclosures. Boone and White [2015] demonstrates that higher institutional ownership is associated more capital market attention and more voluntary disclosures. Inspired by their research, we define *Inst* to measure sophisticated investor attention, which equals 1 if the ownership of institutional investors are above the mean value and 0 if otherwise. We split the sample into two groups based on the value of *Inst* and re-estimate Equation (4) in two sub-samples. The results presented in Panel B of Table 6 show that the positive relations between NORs and analyst forecast errors, dispersion, and uncertainty are present only in firms with lower institutional investor ownership. As shown in the last row, Fisher's permutation test of the difference of the coefficients on NOR$^{Firm}$ between the two sub-samples is statistically significant.

*5.7.2. Information Uncertainty* If information uncertainty is the channel through which NORs affect analyst forecasts, the results from our main regression should be more pronounced among firms with higher operational risks and uncertainties. It is well documented that R&D expenditures can intensity firms' future risks (e.g., Shi [2003], Kothari, Laguerre and Leone [2002]). Therefore, we define H_RD as a measure for firm risk, which equals to 1 if firms' R&D expenses are higher than the average and 0 if otherwise. We report the regression results in Table 7, Panel A. The correlations between NORs and analyst forecast features are positive at 1% significance level for firms with higher R&D investments, in contrast, they are insignificant for firms with lower R&D investments. Fisher's permutation test of the difference of the coefficients on NOR$^{Firm}$ between the two sub-samples is statistically significant.

During the COVID-19 pandemic, firms were faced with greater operational uncertainties. Therefore, we define COVID as another measure for firms' risks, which equals 1 if the earnings calls were held during the period of COVID-19 pandemic, and 0 if otherwise. We identify the start of the COVID-19 period as the first reported case in December 2019 and the end as the WHO's official announcement declaring the pandemic's end in May 2023. As reported in Panel B of Table 7, the coefficients of NORs on analyst forecast features are statistically significant only for firms holding earnings calls during COVID-19 period. The Fisher's permutation test of the difference of the coefficients on NOR$^{Firm}$ between the two sub-samples is statistically significant.

Overall, these cross-sectional analyses validate our hypothesis that NORs affect analyst forecast behaviors by exacerbating information asymmetry and increasing information uncertainty.

**TABLE 7**
*Information Uncertainty*

| Panel A: R&D Expenditure | | | | | | |
|---|---|---|---|---|---|---|
| | Error$^{Firm}$ | | Dispersion$^{Firm}$ | | Uncertainty$^{Firm}$ | |
| | H_RD=1 (1) | H_RD=0 (2) | H_RD=1 (3) | H_RD=0 (4) | H_RD=1 (5) | H_RD=0 (6) |
| NOR$^{Firm}$ | 0.000 | 0.001*** | 0.002 | 0.009*** | 0.002 | 0.008*** |
| | (1.21) | (2.65) | (0.35) | (2.76) | (0.49) | (2.90) |
| Controls | Yes | Yes | Yes | Yes | Yes | Yes |
| Firm | Yes | Yes | Yes | Yes | Yes | Yes |
| Quarter | Yes | Yes | Yes | Yes | Yes | Yes |
| Constant | 0.036 | 0.009 | -0.172 | 0.652* | -0.171 | 0.586* |
| | (1.33) | (0.33) | (-0.24) | (1.71) | (-0.29) | (1.78) |
| Observations | 1133 | 3146 | 1133 | 3146 | 1133 | 3146 |
| Adj. $R^2$ | 0.244 | 0.332 | 0.554 | 0.542 | 0.539 | 0.553 |
| Diff. in coef. | P-Value=0.013** | | P-Value=0.013** | | P-Value=0.012** | |
| Panel B: The Influence of COVID-19 | | | | | | |
| | Error$^{Firm}$ | | Dispersion$^{Firm}$ | | Uncertainty$^{Firm}$ | |
| | COVID=1 (1) | COVID=0 (2) | COVID=1 (3) | COVID=0 (4) | COVID=1 (5) | COVID=0 (6) |
| NOR$^{Firm}$ | 0.001*** | 0.000 | 0.008** | 0.009 | 0.008*** | 0.008 |
| | (3.22) | (1.27) | (2.38) | (1.58) | (2.76) | (1.57) |
| Controls | Yes | Yes | Yes | Yes | Yes | Yes |
| Firm | Yes | Yes | Yes | Yes | Yes | Yes |
| Quarter | Yes | Yes | Yes | Yes | Yes | Yes |
| Constant | 0.024 | -0.018 | 0.433 | -0.166 | 0.399 | -0.184 |
| | (0.77) | (-0.55) | (1.04) | (-0.27) | (1.14) | (-0.35) |
| Observations | 3047 | 1228 | 3047 | 1228 | 3047 | 1228 |
| Adj. $R^2$ | 0.342 | 0.256 | 0.575 | 0.480 | 0.586 | 0.469 |
| Diff. in coef. | P-Value=0.003** | | P-Value=0.033** | | P-Value=0.020** | |

Panel A of the table compares the baseline regression results between firms which are more active in R&D activities (*RD*=1) and firms which are less active in R&D activities (*RD*=0). Due to the substantial missing values of R&D expenditures in firms' quarterly reports in our sample, we measure firms' R&D activities by using the value of intangible assets. We report the p value of the Fisher's Permutation test (two-sided) for the difference in the coefficient on NOR$^{Firm}$ between the two sub-samples in the last row. Panel B of the table compares the baseline regression results between firms that held earnings calls during the COVID-19 period (*COVID* = 1) and those that held them outside the COVID-19 period (*COVID* = 0). We identify the start of the COVID-19 period as the first reported case in December 2019 and the end as the WHO's official announcement declaring the pandemic's end in May 2023. We report the p value of the Fisher's Permutation test (two-sided) for the difference in the coefficient on NOR$^{Firm}$ between the two sub-samples in the last row. For brevity, the coefficients of other control variables are omitted. The t statistics, estimated on robust standard errors are reported in parentheses. *, ** and *** denote statistical significance at the 10%, 5% and 1% levels, respectively.

### 5.8 CHANNEL ANALYSIS

*5.8.1. Information Processing Costs* To investigate how non-responses (NORs) contribute to information asymmetry, we further examine their impact on analysts' information processing costs. We argue that Earnings calls with higher incidence of NORs compel analysts to allocate more of their limited time or attention to gathering alternative supporting information, thereby increasing their information burden. This elevated cost may ultimately lead to poorer forecast performance. Following Barinov, Park and Yıldızhan [2024], we utilize the degree of post earnings announcement drift (PEAD) to proxy for information processing costs. As reported in Panel A of Table 8, the coefficient on NOR$^{Firm}$ × Qr_UEPS are significantly positive, indicating NORs leading to greater PEAD. This result suggests more NORs in earnings calls can increase analyst information processing costs.

*5.8.2. Information Uncertainty* In the next, we directly examine how NORs affect information uncertainty. We use the return volatility (*Ret_Sd*), average trading volume (*Volume*), and average bid-ask spread (*Spread*) within 30 days following the earnings call as the measures for information uncertainty (e.g., Zhang [2006]). The results are reported in Panel B of Table 8. It is shown that NORs is positively associated with return volatility, trading volume, and bid-ask spread, which is consistent with our argument that NORs would increase information uncertainty.

## TABLE 8
*Chanel Analysis*

### Panel A: Post Earnings' Announcement Drift

|  | (1) $CAR_{2-60}$ | (2) $BHAR_{2-60}$ | (3) $CAR_{2-60}$ | (4) $BHAR_{2-60}$ |
|---|---|---|---|---|
| Qr_UEPS | 0.006* | 0.006** | 0.000 | -0.000 |
|  | (1.96) | (2.18) | (0.01) | (-0.01) |
| $NOA^{Firm}$ × Qr_UEPS |  |  | 0.004** | 0.004* |
|  |  |  | (2.06) | (1.91) |
| $NOA^{Firm}$ |  |  | -0.010** | -0.009* |
|  |  |  | (-2.02) | (-1.87) |
| Controls | No | No | Yes | Yes |
| Controls × Qr_UEPS | No | No | Yes | Yes |
| Firm | Yes | Yes | Yes | Yes |
| Quarter | Yes | Yes | Yes | Yes |
| Constant | 1.404*** | 1.405*** | 1.416*** | 1.423*** |
|  | (11.55) | (11.83) | (10.68) | (10.80) |
| Observations | 4090 | 4090 | 4090 | 4090 |
| Adj. $R^2$ | 0.091 | 0.095 | 0.092 | 0.096 |

### Panel B: Information Uncertainty

|  | (1) Ret_Sd | (2) Ret_Sd | (3) Volume | (4) Volume | (5) Spread | (6) Spread |
|---|---|---|---|---|---|---|
| $NOR_{it}^{Firm}$ | 0.000* | 0.000* | 0.011** | 0.010** | 0.000* | 0.000* |
|  | (1.87) | (1.69) | (2.48) | (2.27) | (1.86) | (1.70) |
| Controls | Yes | Yes | Yes | Yes | Yes | Yes |
| Firm | Yes | Yes | Yes | Yes | Yes | Yes |
| Quarter | Yes | Yes | Yes | Yes | Yes | Yes |
| Constant | 0.021*** | -0.003 | 11.706*** | 9.986*** | 0.026*** | 0.009 |
|  | (105.65) | (-0.29) | (1472.83) | (18.89) | (146.26) | (0.89) |
| Observations | 4052 | 4052 | 4052 | 4052 | 4052 | 4052 |
| Adj. $R^2$ | 0.580 | 0.592 | 0.912 | 0.920 | 0.652 | 0.667 |

Panel A of the table presents the empirical results examining the effect of NORs on Post Earnings' Announcement Drift (PEAD). Greater PEAD indicates higher information processing costs. The regression model is as follows:

$$PEAD_{it} = \beta_0 + \beta_1 Qr\_UEPS_{it} + \beta_2 Qr\_UEPS_{it} \times NOA_{it}^{Firm} + \beta_3 NOA_{it}^{Firm} + \beta_4 Controls_{it} + \beta_5 Qr\_UEPS_{it} \times Controls_{it} + \sum Firm + \sum Quarter + \varepsilon_{it}$$

We use market-adjusted cumulative abnormal return ($CAR_{2-60}$) and buy-and-hold abnormal return ($BHAR_{2-60}$) over the earnings window [2-60] to compute PEAD. We add control variables including market value (*Mkv*), return on assets (*Roa*), book-to-market ratio (*BM*), whether earnings are negative (*Loss*), analyst following (*Num_Ana*), and equal-weighted returns (*Wt_Ret*). Panel B of the table presents the empirical results examining the effect of NORs on information uncertainty in the capital market. We use stock return volatility within the 30 days following the earnings call (*Ret_Sd*), trading volume within the 30 days following the earnings call (*Volume*), and bid-ask spread within the 30 days following the earnings call (*Spread*) to proxy for information uncertainty. All control variables and their productions with *Qr_UEPS* are omitted for brevity. The detailed definitions of the variables are provided in Appendix B. The t statistics, estimated on robust standard errors are reported in parentheses. *, ** and *** denote statistical significance at the 10%, 5% and 1% levels, respectively.

## 6. Conclusion

Despite the prevalence of managers' non-responses during earnings calls, few studies examine their impact on the analyst forecast behaviors—despite analysts being the direct direct interlocutors of management in these Q&A sessions. Taking advantage of the most-advanced large language models, we develop a novel measure for managers' non-responses. Specifically, we prompt the most representative close-source and open-source models–Chat-GPT4 and Llama3.3–to perform three tasks, namely, identification, classification, and evaluation.

We compare the performances of the two models and find that Chat-GPT4 identifies systematically more non-responses than Llama3.3. Nevertheless, the distributions and year-trend patterns of NORs remain similar across the two models. We then construct the measure for non-response (NORs) with the identifications from Chat-GPT4 and explore how NORs affect analyst forecast features. It is found that (Chat-GPT4 identified) NORs in earnings calls are positively related to analyst forecast errors, dispersion, and forecast uncertainty. The baseline regression result is robust with alternative measures and controlling managerial incentives. We propose and demonstrate that information asymmetry and uncertainty are the two channels through which NORs affect analyst forecast features.

We also investigate the heterogeneous effects of various NOR types on analyst forecast characteristics, based on Chat-GPT4's classifications of managers' underlying reasons for providing a non-response. The result suggests that "direct refusal" dominates the relation between NORs and analyst forecast features. Additionally, we re-estimate the main regression using Chat-GPT4's evaluations of managers' responses along the three dimensions of Grice's Maxims—quantity, clarity, and relevance. Among these, only the relevance score exhibits a statistically significant negative association with analyst forecast characteristics.

In a nutshell, our research highlights to the regulators the potential adverse impact of managers' non-responses in the earnings calls. Distinguishing the motivations behind NORs is crucial to curbing such kinds of phenomena.

This paper illustrates the application of large language models in developing measures of interest, facilitating research in accounting. Our approach and prompt can be applied to many other public communication settings involving Q&A interactions, such as online investor interaction platforms, social media discussions, and public statements by politicians or government officials. As a caveat, researchers should be careful when leveraging LLMs to assist research, as their responses can occasionally be misleading or deceptive. While our research shows that well-crafted prompt can help alleviate this problem to some degree, a comprehensive post-hoc manual checking is also necessary. Moreover, data reproducibility still remains a matter of concern. In this paper, we employ multiple models and apply manually checking, bootstrapping and resampling techniques to mitigate the issue as much as possible. We suggest studies in the future to explore more promising resolutions to these problems.

# APPENDIX A: PROMPT TO ELICIT MANAGERS' NON-RESPONSES IN EARNINGS CALLS

The following boxes demonstrate the prompting process to elicit managers' non-responses(NORs) in firms' earnings call transcripts. We adopt a few-shot strategy with two examples provided. Figure 1 to Figure 2 present the prompt format including system message[23] and user prompts. Figure 3 to Figure 4 show the two examples.

---
23. The system prompt is "You are a helpful research assistant in accounting and finance."

FIG. 1 Prompt Template

```
User Prompt_Template="""
```
The provided document is a conversation between an analyst and a manager during a company's earnings call.
The types of speakers are shown in square brackets as ["Speaker Type"], followed by the content of the speech of corresponding speakers.
You have three tasks to finish:
Task 1 - Identify Non-Response Q&A Exchanges:

- Extract Q&A exchanges where managers explicitly refused to answer the analyst's question, using responses such as "I can't answer" or "I don't know."

- Extract Q&A exchanges where managers provided answers that were irrelevant and did not address the analyst's question.

- Exclude Q&A exchanges where managers sought clarification of the question or got disconnected.

Task 2 - Classify the non-responses. If any non-responses were identified in Task 1, classify it into one of the five categories:

- Refusal: if managers directly refuse to answer the question without giving any justifications.

- Lack of Info: if managers indicate their lack of relevant information to answer the question.

- Legal Affairs: managers can't provide answers due to legal restrictions.

- Recall: if managers indicate they will get back to the question sometime in the future.

- Irrelevant: if managers give irrelevant answers.

- If you didn't identify any non-responses in Task 1, return null for this task.

Task 3 - Evaluate and rate managers' overall responses on a scale of 0-10 from three aspects: quantity, relevance, and clarity. Please note 0 means least informative, relevant and clear, while 10 indicates most informative, relevant and clear.

- Quantity: rate managers' responses according to the amount of incremental information provided in their answers.

- Relevance: rate managers' responses based on the relevance of their answers to the analyst's questions.

- Clarity: rate managers' responses based on the clarity of their answers.

```
 After finishing the two tasks, output your response in following JSON
format: {
"NOR":  1, If any Non-Response Q&A Exchanges were identified; 0, if no
Non-Response Q&A Exchanges were identified,
"Pair":  [Store both the question and the answer in the list if any
Non-response Q&A Exchanges were identified,or write null, if not any pairs
identified]
"Category":  Store the result from Task 2
"Quantity":  Score of quantity,
"Relevance":  Score of relevance,
"Clarity": Score of clarity
}
Here are two examples for your references:
[Example 1]:
Statement=statement 1
Responses = response 1
[Example 2]:
Statement=statement 2
Responses = response 2
**Important Rules:  **

   • Strictly adhere to the output formats as provided above.

   • For every transcript, there should be only one library being output.

   • If there are more than one pair of exchanges which have no responses,
     store all of them as value for "Pair" separated by "," in the list.

   • For Task 3, you should generate the scores based on the managers'
     overall responses.

Begin the analysis now.
Statement=statement
Responses =
""".strip()
```

FIG. 3 Demonstrations

```
 Statement 1="""
......
[Analyst]:"And the Dubai client, are they a financially distressed client?
Just kind of characterize how you feel about being able to ultimately
collect from them."
[Manager]:"Yes. No.  So we have no knowledge that the client has any
financial issue or is unable to pay. The reserve was really a function
of the aging of the receivables and the inability for us, along with the
client to agree upon a revised payment plan. And really, at that point,
we felt it was appropriate to take the reserve.  But it has nothing to do
with any knowledge or understanding of an inability to pay." """

Response 1="""{
"NOR": 1,
"Pair": [{
[Analyst]:"And the Dubai client, are they a financially distressed client?
Just kind of characterize how you feel about being able to ultimately
collect from them.",
[Manager]:"Yes. No.  So we have no knowledge that the client has any
financial issue or is unable to pay. The reserve was really a function
of the aging of the receivables and the inability for us, along with the
client to agree upon a revised payment plan. And really, at that point,
we felt it was appropriate to take the reserve.  But it has nothing to do
with any knowledge or understanding of an inability to pay." },]
"Quantity": Lack of Info,
"Quantity": 3,
"Relevance":  10,
"Clarity": 10
}"""
```

FIG. 4 Demonstrations Continued

```
 Statement 2="""
......
[Analysts][James Salera]:"You've already discussed some of the trends
with Quest in the category.  But I was wondering if you could give us a
sense for how much of the broader category growth is being driven by the
expansion of products and the appeal that, that brings to expand buy rates
versus consumers that are increasingly health conscious kind of engaging
with these protein dense cohorts, low-calorie snacks?"
[Executives][Geoff Tanner]:"Yes. No, it's a good question.  I don't
have that information at a category level.  As we look at our brands,
we certainly see a balance across both household penetration and buy rate.
And as I said, again, to Matt's – with Matt's question, this category
largely grew up as bars and shakes. And over time, has expanded well
beyond that new format, new usage occasions, new dayparts."
......
"""

Response 2="""{
"NOR": 1,
"Pair":[{
[Analysts][James Salera]:"You've already discussed some of the trends
with Quest in the category.  But I was wondering if you could give us a
sense for how much of the broader category growth is being driven by the
expansion of products and the appeal that, that brings to expand buy rates
versus consumers that are increasingly health conscious kind of engaging
with these protein dense cohorts, low-calorie snacks?",
[Executives][Geoff Tanner]:"Yes. No, it's a good question.  I don't
have that information at a category level.  As we look at our brands,
we certainly see a balance across both household penetration and buy rate.
And as I said, again, to Matt's – with Matt's question, this category
largely grew up as bars and shakes. And over time, has expanded well
beyond that new format, new usage occasions, new dayparts." },]
"Quantity": Lack of Info,
"Quantity": 4,
"Relevance":  9,
"Clarity": 8
}"""
```

# APPENDIX B: VARIABLE DEFINITIONS

**Call-level Variables**

| | |
|---|---|
| $NOR_C$ | Indicator variable equal to 1 at the conversation level if any non-responses occur during Q&A exchanges between an analyst and management. |
| $NOR^{Con}$ | Total number of non-responses during Q&A exchanges between an analyst and management. |
| $NOR_F$ | Indicator variable equal to 1 if any non-responses occur during the earnings call. |
| $NOR^{Firm}$ | Total number of non-responses during the earnings call. |
| Quantity | LLMs' evaluation of the incremental information content in the manager's responses. |
| Relevance | LLMs' evaluation of the relevance of the manager's responses. |
| Clarity | LLMs' evaluation of the clarity of the manager's responses. |
| Mscore | The average value of LLMs' evaluations of managers' responses in terms of *Quantity*, *Relevance*, and *Clarity*. |
| $Error^{Firm}$ | Absolute difference between the I/B/E/S reported quarterly EPS and the most recent analyst consensus after the earnings announcement, divided by the closing price of the previous financial quarter. |
| $Dispersion^{Firm}$ | Standard deviation of individual analyst forecasts in the most recent post-call I/B/E/S summary, divided by the closing price of the previous financial quarter. |
| $Uncertainty^{Firm}$ | Referring to Lehavy, Li and Merkley [2011], we interpret uncertainty as the sum of idiosyncratic and common uncertainty. Idiosyncratic uncertainty reflects the uncertainty associated with analysts' private information and common uncertainty reflects the uncertainty associated with the information common to all analysts. It is computed as the following equation: $$Uncertainty_{it} = \left(1 - \frac{1}{Analyst\ Following_{it}}\right) * Dispersion_{it} + Error_{it} \quad (5)$$ |
| SurEar | Earnings surprise, calculated as the EPS of quarter $t$ minus that of quarter $t-1$, divided by the closing price of quarter $t-1$. |
| Size | Firm size, computed as the logarithm of total assets plus 1. |
| Roa | Return on assets, calculated as income before extraordinary items divided by total assets. |
| RetVol | Return volatility, measured as the standard deviation of stock returns over the prior 12 months. |
| Loss | Indicator variable equal to 1 for negative net earnings, 0 otherwise. |
| *Mkv* | Logarithm of market value plus 1. |
| *Lev* | Financial leverage, calculated as total debt divided by total assets. |
| *Tone* | Difference between the frequency of positive and negative words, divided by their sum in managers' presentation: |

$$Tone_{it} = \frac{Positive\ Words_{it} - Negative\ Words_{it}}{Positive\ Words_{it} + Negative\ Words_{it}}$$

where *Positive Words*$_{it}$ and *Negative Words*$_{it}$ denote the respective word counts in the manager's presentation during the earnings call for firm *i* in quarter *t*.

| | |
|---|---|
| *Uncert* | Ratio of uncertainty-related words to total words in the presentation, computed as: |

$$Uncert_{it} = \frac{Uncertainty\ Words_{it}}{Total\ Words_{it}}$$

where *Uncertainty Words*$_{it}$ is the number of uncertainty-related words and *Total Words*$_{it}$ is the total word count in the manager's presentation for firm *i* in quarter *t*. The uncertainty word list is sourced from Loughran and McDonald [2014].

| | |
|---|---|
| *Forward* | The ratio of forward-looking terms to the total word count in the presentation, computed as the follows: |

$$Forward_{it} = \frac{Forward\_looking\ Words_{it}}{Total\ Words_{it}}$$

Where *Forward_looking Words*$_{it}$ represents the number of forward-looking words in the manager's presentation during earnings call at quarter *t* for firm *i*. *Total Words*$_{it}$ represents total words in the manager's presentation during earnings call at quarter *t* for firm *i*. Forward_looking word list comes from Bozanic, Roulstone and Van Buskirk [2018].

| | |
|---|---|
| *Read* | Fog index, referring to the methodologies of Li [2008] and Loughran and McDonald [2014]. Computed as follows: |

$$Fog\ Index_{it} = 0.4 \times \left(\frac{Total\ Words_{it}}{Sentences_{it}}\right) + 100 \times \left(\frac{Complex\ Words_{it}}{Total\ Words_{it}}\right)$$

Where *Sentences*$_{it}$ represents the number of sentences in the manager's presentation during the earnings call at quarter *t* for firm *i*. *Complex Words*$_{it}$ represents the number of complex words in the manager's presentation during the earnings call at quarter *t* for firm *i*. *Total Words*$_{it}$ represents the total number of words in the manager's presentation during the earnings call at quarter *t* for firm *i*. Complex words are words with more than 3 syllables, but are not compound words.

| | |
|---|---|
| *MO* | An indicator variable that equals 1 if the firm's operation complexity is above the sample mean value, and 0 otherwis. Following Barinov, Park and Yıldızhan [2024], we measure firms' operation complexity by counting the number of firms' divisions with different two-digit SIC codes. |
| *Inst* | An indicator variable that equals 1 if the institutional ownership is above the sample mean value, and 0 otherwise. We calculate institutional ownership as the total percentage of outstanding shares held by institutional investors[24]. |
| *H_RD* | An indicator variable that equals 1 if the value of intangible assets is above the sample mean value, and 0 otherwise. |
| *COVID* | An indicator variable that equals 1 if the calendar quarter of earnings calls conference is between 2019Q4 (the first official report of COVID-19) and 2023Q2 (the official end of COVID-19 declared by WHO). |
| *Ret_Sd* | Stock return volatility within the 30 days following the earnings call. |
| *Volume* | Trading volume within the 30 days following the earnings call. |
| *Spread* | The mean bid-ask spread within 30 days following the earnings call. |
| *CAR* | The market-adjusted cumulative abnormal stock return over the earnings window [2-60]. |
| *BHAR* | The cumulative buy-hold abnormal stock return over the earnings window [2-60]. |
| *Rd_Exp* | R&D expenditures, computed as the R&D expenses reported in the financial statements divided by operation expenses. [25] |
| *BM* | Book-to-market ratio, computed as follows: $$\frac{\text{Book Value of Common Shares}_{it}}{\text{Stock Price}_{it} \times \text{The Number of Common Shares Outstanding}_{it}}$$ The Stock Price$_{it}$ is the average of the bid and ask price during the last month of quarter t for firm i. |
| *Num_Ana* | Analyst following, computed as the logarithm of one plus the number of analysts issuing EPS estimates for a firm. The information of analyst estimates is from the summary history document of I/B/E/S. |
| *We_Ret* | Equal-weighted returns is the ewretd from CRSP/Compustat merged database - fundamentals quarterly. |
| **Conversation-level Variables** | |
| *Word* | The number of words of the conversation between an analyst and the management |
| *Order* | The order of the Q&A exchange occurs in an earnings call. |
| *Tone_Q* | Tone computed at conversation level, similar to Tone metric. |

---

24. it is noted as "InstOwn_Perc" in Thomson Reuters Institutional Holdings (13F form) database.
25. Following Lehavy, Li and Merkley [2011], we replace all the missing values with 0.

| | |
|---|---|
| *Uncert_Q* | Uncertain information disclosure computed at conversation level, similar to Uncert metric. |
| *Forward_Q* | Forward-looking information disclosures computed at conversation level, similar to Forward metric. |
| *Read_Q* | Fog Index computed at the conversation level, similar to the readability metric. |

**Individual-level Variables**

| | |
|---|---|
| Time$^{Individual}$ | Natural logarithm of the number of days between the earnings announcement and the most recent EPS forecast/revision within 30 days, plus 1. |
| Error$^{Individual}$ | Absolute difference between the individual analyst EPS forecast (within 30 days after the earnings announcement) and the actual EPS, divided by the closing price of the previous financial quarter. |
| *Comp* | Stock price-based compensation. Following Hollander, Pronk and Roelofsen [2010], it is computed as the ratio of the average annual ex ante stock price-based compensation (i.e., the sum of the total value of stock-option grants plus the value of restricted stock grants) to the total direct managerial compensation. |
| *Lwealth* | Stock price-based wealth. Following Hollander, Pronk and Roelofsen [2010], it is computed as the natural logarithm of the value of managerial shareholdings in the firm (using the fiscal year-end closing price). |

## APPENDIX C: SAMPLE SELECTION

| Selections | Transcripts | Conversations |
|---|---|---|
| S&P 500 Earnings Calls Transcripts from Capital-IQ | 10,976 | 107,564 |
| less duplicate conversations | (23) | (307) |
| | 10,953 | 107,257 |
| less transcripts without Chat-GPT4's responses | (338) | (416) |
| | 10,615 | 106,841 |
| less transcripts without gvkey linking to Compustat | (240) | (2,337) |
| | 10,375 | 104,504 |
| less transcripts without permno linking to CRSP | (2,896) | (29,140) |
| | 7,479 | 75,364 |
| less transcripts without analyst forecasts from I/B/E/S | (2,565) | (23,563) |
| | 4,583 | 51,801 |
| less transcripts without financial fundamentals from Compustat | (96) | (865) |
| | 4,487 | 50,936 |
| less transcripts without stock market indicators from CRSP | (203) | (3,102) |
| | 4,284 | 47,834 |
| **Final Sample** | **4,284** | **47,834** |

This table shows sample selection process. We only present the major steps. The number shown in the parentheses are the dropped sample number at each step.

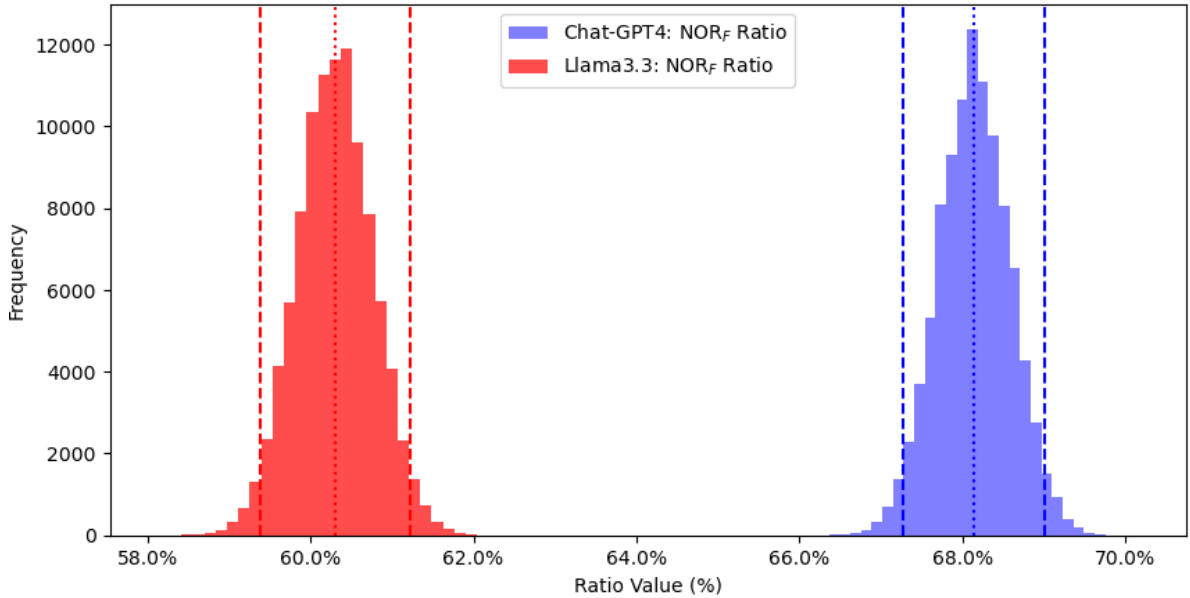

FIG. 1 — Sample distributions of bootstrap for earnings' calls.

This figure visualizes the distributions of bootstrapping results for quarterly NOR percentages at the call level ($NOR_F$ Ratio) for the whole sample of transcripts. The histogram in blue represents Chat-GPT4's identifications and the histogram in red represents Llama3.3's identifications. $NOR_F$ Ratio is computed the number of earnings calls with NORs divided by the total number of earnings calls. The bootstrapping iterates 100,000 times.

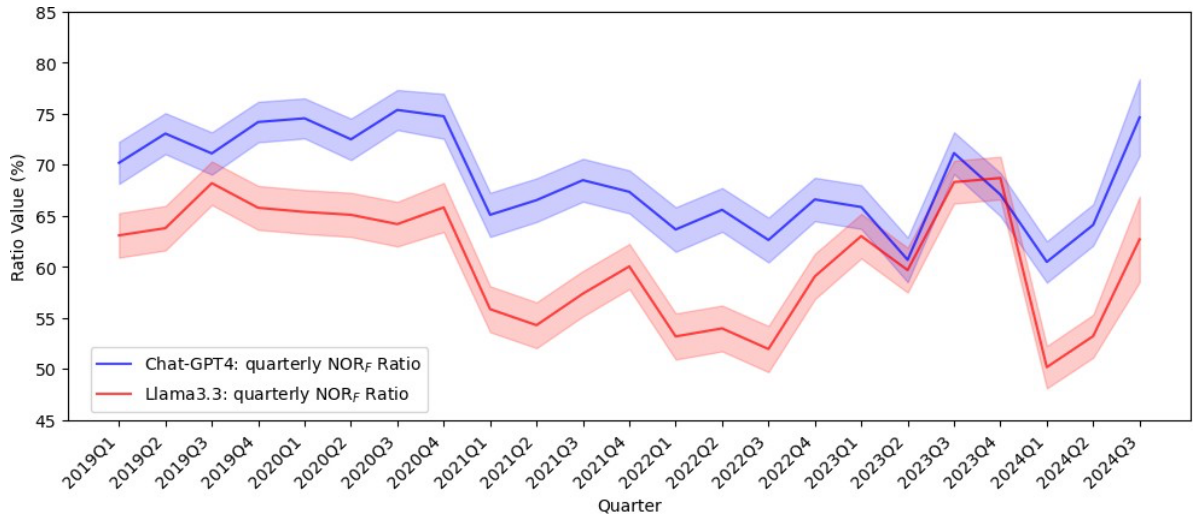

FIG. 2 — Call Level: Time Series Plot with Bootstrapped Confidence Intervals.

This figure displays the trend of quarterly NOR percentage at the call level ($NOR_F$ Ratio) with bootstrapped confidence intervals raging from 2019Q1 to 2024Q3. The plot in blue represents Chat-GPT4's identifications and the plot in red represents Llama3.3's identifications. $NOR_F$ Ratio is computed as the number of earnings calls with NORs divided by the total number of earnings calls in a quarter. The line is the actual value of quarterly NOR ratio. The shaded areas represent one standard deviation above and below the actual value. The bootstrapping iterates 100,000 times.

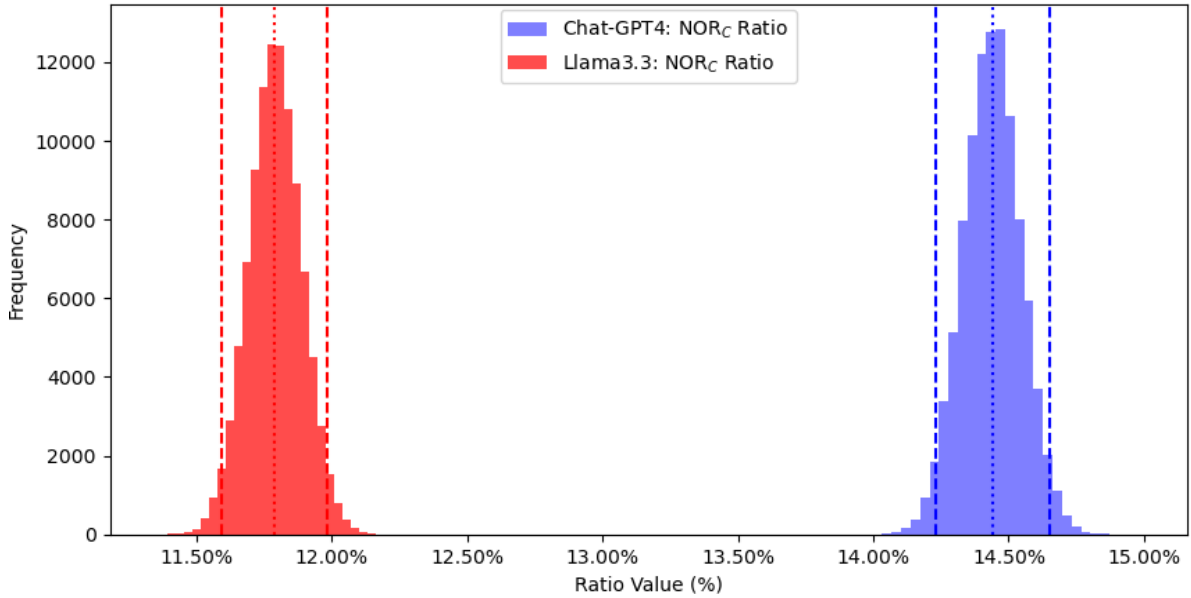

FIG. 3 — Sample Distributions of Bootstrap for Conversations.

This figure visualizes the distributions of bootstrapping results for NOR percentages at the conversation level ($NOR_C$ Ratio) for the whole sample of transcripts. The histogram in blue represents Chat-GPT4's identifications and the histogram in red represents Llama3.3's identifications. $NOR_C$ Ratio is computed the number of conversations with NORs divided by the total number of conversations from all the earnings calls. The bootstrapping iterates 100,000 times.

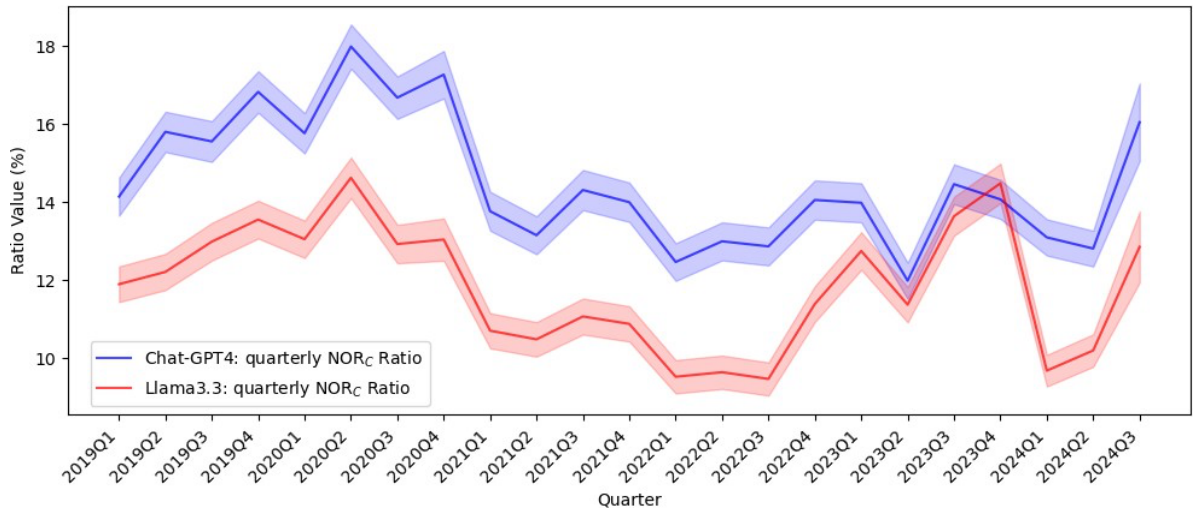

FIG. 4 — Conversation Level: Time Series Plot with Bootstrapped Confidence Intervals.

This figure displays the trend of quarterly NOR percentage at the conversation level ($NOR_C$ Ratio) with bootstrapped confidence intervals raging from 2019Q1 to 2024Q3. The plot in blue represents Chat-GPT4's identifications and the plot in red represents Llama3.3's identifications. $NOR_C$ Ratio is computed as the number of conversations with NORs divided by the total number of conversations from all the earnings calls in a quarter. The line is the actual value of quarterly NOR ratio. The shaded areas represent one standard deviation above and below the actual value. The bootstrapping iterates 100,000 times.

**TABLE 1**
*Non-responses Statistics by Quarter*

| (1) Quarter | (2) NOR$_F$ | (3) NOR$_C$ | (4) Refusal | (5) Recall | (5) Lack | (6) Irrelevant | (7) Legal | (8) Others |
|---|---|---|---|---|---|---|---|---|
| Panel A: Responses from Chat-GPT4 | | | | | | | | |
| 2019q1 | 70.18% | 14.34% | 4.98% | 2.18% | 5.81% | 1.11% | 0.42% | 0.00% |
| 2019q2 | 73.05% | 16.03% | 5.25% | 2.64% | 6.21% | 1.56% | 0.52% | 0.02% |
| 2019q3 | 71.10% | 15.71% | 5.00% | 2.48% | 6.34% | 1.65% | 0.41% | 0.00% |
| 2019q4 | 74.18% | 16.98% | 5.10% | 3.04% | 7.09% | 1.33% | 0.64% | 0.02% |
| 2020q1 | 74.54% | 15.90% | 5.06% | 2.33% | 6.98% | 1.46% | 0.28% | 0.00% |
| 2020q2 | 72.48% | 18.27% | 3.91% | 2.18% | 10.92% | 1.29% | 0.22% | 0.00% |
| 2020q3 | 75.36% | 16.89% | 4.59% | 2.90% | 7.96% | 1.39% | 0.32% | 0.02% |
| 2020q4 | 74.74% | 17.48% | 4.72% | 3.10% | 8.23% | 1.29% | 0.39% | 0.03% |
| 2021q1 | 65.09% | 13.93% | 3.79% | 2.13% | 6.53% | 1.26% | 0.36% | 0.02% |
| 2021q2 | 66.53% | 13.34% | 3.82% | 2.06% | 5.88% | 1.39% | 0.33% | 0.02% |
| 2021q3 | 68.48% | 14.57% | 3.90% | 2.16% | 7.04% | 1.32% | 0.37% | 0.00% |
| 2021q4 | 67.34% | 14.15% | 3.81% | 2.54% | 6.24% | 1.40% | 0.36% | 0.00% |
| 2022q1 | 63.66% | 12.63% | 3.78% | 1.95% | 5.33% | 1.53% | 0.21% | 0.00% |
| 2022q2 | 65.58% | 13.28% | 3.57% | 2.16% | 5.83% | 1.43% | 0.44% | 0.00% |
| 2022q3 | 62.63% | 12.93% | 3.39% | 2.10% | 5.45% | 1.76% | 0.36% | 0.00% |
| 2022q4 | 66.60% | 14.20% | 3.62% | 2.66% | 6.29% | 1.51% | 0.29% | 0.00% |
| 2023q1 | 66.15% | 13.93% | 4.10% | 2.05% | 6.26% | 1.43% | 0.34% | 0.00% |
| 2023q2 | 60.69% | 12.02% | 3.19% | 1.80% | 5.13% | 1.63% | 0.33% | 0.00% |
| 2023q3 | 71.14% | 14.56% | 4.48% | 2.58% | 5.63% | 1.68% | 0.29% | 0.00% |
| 2023q4 | 67.07% | 14.10% | 3.87% | 2.78% | 5.42% | 1.61% | 0.47% | 0.02% |
| 2024q1 | 60.55% | 13.29% | 3.77% | 2.49% | 5.25% | 1.46% | 0.45% | 0.00% |
| 2024q2 | 64.29% | 13.07% | 3.57% | 1.95% | 5.54% | 1.75% | 0.38% | 0.00% |
| 2024q3 | 74.63% | 16.57% | 4.62% | 2.64% | 6.52% | 2.57% | 0.51% | 0.00% |
| Panel B: Responses from Llama3.3 | | | | | | | | |
| 2019q1 | 63.29% | 11.94% | 2.97% | 0.50% | 7.39% | 1.84% | 0.12% | 0.57% |
| 2019q2 | 64.07% | 12.22% | 3.09% | 0.52% | 7.74% | 1.95% | 0.10% | 0.50% |
| 2019q3 | 68.40% | 13.04% | 3.12% | 0.47% | 8.75% | 1.91% | 0.08% | 0.47% |
| 2019q4 | 66.19% | 13.68% | 3.58% | 0.84% | 9.07% | 1.73% | 0.20% | 0.36% |
| 2020q1 | 65.58% | 13.16% | 3.16% | 0.54% | 8.87% | 1.85% | 0.08% | 0.42% |
| 2020q2 | 66.12% | 14.75% | 2.29% | 0.39% | 11.12% | 1.62% | 0.04% | 0.67% |
| 2020q3 | 64.21% | 12.90% | 2.53% | 0.55% | 9.53% | 1.35% | 0.06% | 0.29% |
| 2020q4 | 66.26% | 13.26% | 2.81% | 0.92% | 8.97% | 1.69% | 0.06% | 0.42% |
| 2021q1 | 55.94% | 10.76% | 2.08% | 0.44% | 7.39% | 1.49% | 0.08% | 0.38% |
| 2021q2 | 54.49% | 10.54% | 2.07% | 0.56% | 7.09% | 1.43% | 0.02% | 0.56% |
| 2021q3 | 57.37% | 11.06% | 2.25% | 0.48% | 7.61% | 1.60% | 0.09% | 0.28% |
| 2021q4 | 60.45% | 10.93% | 2.19% | 0.82% | 7.19% | 1.39% | 0.04% | 0.57% |
| 2022q1 | 53.39% | 9.56% | 2.02% | 0.40% | 6.53% | 1.49% | 0.04% | 0.36% |
| 2022q2 | 53.97% | 9.67% | 1.75% | 0.40% | 6.68% | 1.37% | 0.17% | 0.23% |
| 2022q3 | 52.16% | 9.50% | 1.88% | 0.34% | 6.39% | 1.38% | 0.04% | 0.30% |
| 2022q4 | 59.47% | 11.37% | 2.53% | 0.88% | 7.24% | 1.46% | 0.10% | 0.48% |
| 2023q1 | 63.01% | 12.77% | 2.80% | 0.59% | 7.67% | 1.67% | 0.23% | 0.44% |
| 2023q2 | 59.76% | 11.38% | 2.63% | 0.39% | 6.42% | 2.14% | 0.10% | 0.00% |
| 2023q3 | 68.29% | 13.70% | 3.32% | 0.67% | 7.57% | 2.52% | 0.23% | 0.00% |
| 2023q4 | 68.90% | 14.51% | 2.98% | 1.09% | 8.30% | 2.40% | 0.31% | 0.00% |
| 2024q1 | 50.26% | 9.64% | 2.15% | 0.54% | 6.08% | 1.65% | 0.15% | 0.28% |
| 2024q2 | 53.57% | 10.27% | 1.94% | 0.43% | 6.28% | 1.88% | 0.04% | 0.56% |
| 2024q3 | 62.69% | 13.05% | 2.95% | 0.74% | 7.82% | 2.51% | 0.37% | 0.66% |

This table presents the percentage descriptions of NORs by quarter on call (NOR$_F$) and conversation level (NOR$_C$) from Chat-GPT4 (*PanelA*) and Llama3.3 (*PanelB*), respectively. The percentage of NORs is computed as the number of earnings calls or conversations with NORs divided by the total number of calls or conversations during a quarter for all the S&P 500 firms in our sample. Column (3) through (8) provide a detailed description regarding the categories of NORs on the conversation level.

**TABLE 2**
*Match Ratios of Repetitions*

| Panel A: ChatGPT Match Ratios | | | | | |
|---|---|---|---|---|---|
| baseline_NOR | N | Mean | SD | Min | Max |
| 0 | 62 | 87.15 | 30.11 | 0.00 | 100.00 |
| 1 | 38 | 58.39 | 43.02 | 0.00 | 100.00 |
| Total | 100 | 76.22 | 38.04 | 0.00 | 100.00 |
| Panel B: Llama Match Ratios | | | | | |
| baseline_NOR | N | Mean | SD | Min | Max |
| 0 | 50 | 99.98 | 0.14 | 99.00 | 100.00 |
| 1 | 50 | 38.24 | 45.76 | 0.00 | 100.00 |
| Total | 100 | 69.11 | 44.71 | 0.00 | 100.00 |

This table reports summary statistics for the match ratios generated from 100 iterations of sampling with GPT4 and Llama3.3 separately. The baseline sample consists of 100 conversations, with 50% (38%) identified as containing non-answers and the remaining 50%(62%) as not containing non-answers by Llama3.3 (Chat-GPT4). *Match Ratio* is computed as the percentage of iterations that match the identifications of each conversation from the baseline sample. For example, a 100% match ratio means that all iterations agreed with the identifications of the baseline sample.

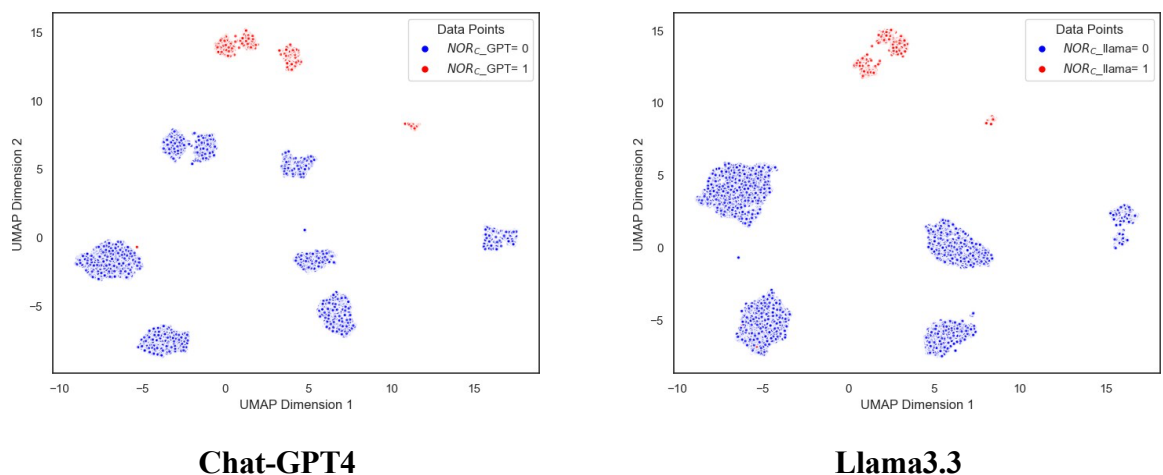

**Chat-GPT4**  **Llama3.3**

FIG. 5 — Clusters of Non-responses

This figure presents the clusters of NORs from Chat-GPT4 (NOR_GPT on the left) and Llama3.3 (NOR_Llama on the right) separately. The clusters are generated using UMAP by reducing potential 24 influencing factors of NORs into two dimensions. These influencing factors include firm-level features: unexpected earnings, firm size, market value, firm leverage, stock return volatility, earnings sale growth, return on assets, loss, R&D expenditures, and the ratio of intangible assets; and conversation-level features: the number of words in a Q&A section, the relative order of Q&A exchanges occure, the models' evaluations on quantity, relevance, and clarity, the tone, uncertainty, forward-information disclosure, and the fog index of analysts' questions.

**TABLE 3**
*Motivations of Non-responses*

|  | Logit | | OLS | |
|---|---|---|---|---|
|  | (1) $GPT\_NOR_{it}^{Con}$ | (2) $Llama\_GPT\_NOR_{it}^{Con}$ | (3) $GPT\_NOR_{it}^{Con}$ | (4) $Llama\_GPT\_NOR_{it}^{Con}$ |
| Word | -0.000*** | -0.001*** | -0.000*** | -0.000*** |
|  | (-6.78) | (-14.65) | (-2.85) | (-8.24) |
| Order | 0.047*** | 0.044*** | 0.006*** | 0.005*** |
|  | (13.02) | (11.65) | (7.81) | (7.37) |
| Quantity | -0.887*** | -0.770*** | -0.125*** | -0.079*** |
|  | (-16.09) | (-12.88) | (-12.89) | (-9.86) |
| Relevance | -1.111*** | -0.704*** | -0.143*** | -0.065*** |
|  | (-25.59) | (-14.95) | (-22.86) | (-12.79) |
| Clarity | 0.299*** | -0.419*** | 0.022*** | -0.092*** |
|  | (5.44) | (-7.63) | (3.23) | (-9.28) |
| Tone_Q | -0.128*** | -0.165*** | -0.013*** | -0.014*** |
|  | (-6.48) | (-7.82) | (-4.98) | (-7.91) |
| Forward_Q | 6.460*** | 9.313*** | 0.728*** | 0.889*** |
|  | (3.99) | (5.53) | (2.85) | (4.85) |
| Read_Q | 0.013*** | 0.018*** | 0.001*** | 0.001*** |
|  | (4.58) | (6.15) | (3.06) | (3.67) |
| Uncert_Q | -1.544 | -0.885 | -0.161 | -0.108 |
|  | (-1.47) | (-0.80) | (-1.42) | (-1.49) |
| Rd_Exp | -0.094 | -0.199 | -0.031 | -0.052 |
|  | (-0.67) | (-1.34) | (-1.53) | (-1.61) |
| Size | 0.002 | 0.026** | 0.015* | 0.012 |
|  | (0.21) | (2.28) | (1.95) | (1.42) |
| Roa | -0.751 | -0.127 | 0.081 | 0.010 |
|  | (-0.96) | (-0.15) | (1.05) | (0.10) |
| Loss | 0.075 | 0.125** | 0.005 | -0.003 |
|  | (1.33) | (2.08) | (0.90) | (-0.48) |
| Firm | Yes | Yes | Yes | Yes |
| Quarter | Yes | Yes | Yes | Yes |
| Constant | 12.319*** | 13.803*** | 2.068*** | 2.032*** |
|  | (29.39) | (32.36) | (16.19) | (17.12) |
| Observations | 51513 | 51513 | 51513 | 51513 |
| Pse./Adj. $R^2$ |  |  | 0.113 | 0.090 |

This table presents the empirical results of adopting a logit model and an ordinary least squares (OLS) model to examine the factors predicting managers' non-responses at the call level. The estimation is based on the following model:

$$Logit/OLS(NOR_{it}) = \alpha_0 + \alpha_1 Word_{it} + \alpha_2 Order_{it} \alpha_3 Quantity_{it} + \alpha_4 Relevance_{it} + \alpha_5 Clarity_{it}$$
$$+ \alpha_6 Tone\_Q_{it} + \alpha_7 Forward\_Q_{it} + \alpha_8 Read\_Q_{it} + \alpha_9 Rd\_Exp_{it} + \alpha_{10} Size_{it} \quad (1)$$
$$+ \alpha_{11} Roa_{it} + \alpha_{12} Loss_{it} + \sum Firm + \sum Quarter + \varepsilon_{it}.$$

Column(1) through (2) present the results from logit model with pseudo $R^2$ reported. Column(3) through (4) present the results from OLS regression with adjusted $R^2$ reported. Column (1) and (3) are estimated on the responses from Chat-GPT4 ($GPT\_NOR_{it}^{Con}$). Column (2) and (4) are estimated on the responses from Llama3.3 ($Llama\_GPT\_NOR_{it}^{Con}$). The control variables include the average values of LLMs' evaluations for quantity (Quantity$_{it}$), relevance (Relevance$_{it}$) and clarity (Clarity$_{it}$) of the manager' answers, the textual features of analysts' questions (Tone_Q$_{it}$, Forward_Q$_{it}$, andFog_Q$_{it}$), R&D expenditure (Rd_Exp$_{it}$), firm size (Size$_{it}$), return on assets (Roa$_{it}$), and firms' earnings performances (Loss$_{it}$). In the OLS model, firm and quarter fixed effects are further included. The detailed definitions of the variables are provided in Appendix B. The t statistics, estimated on robust standard errors are reported in parentheses. *, ** and *** denote statistical significance at the 10%, 5% and 1% levels, respectively.